\documentclass[a4paper]{article}
\usepackage{main/arxiv}
\usepackage{multicol}
\usepackage{todonotes}
\usepackage[normalem]{ulem}
\usepackage{multirow}
\usepackage{url}

\newcommand{\new}[1]{\noindent #1}
\newcommand{\newnew}[1]{\noindent #1\normalcolor}

\usepackage{hyperref}
\usepackage{xr-hyper}
\usepackage{amsmath}
\usepackage{biblatex}
\begin{document}
\twocolumn[

\title{Knitting Multistability}

\author{
 Kausalya Mahadevan \\
  John A. Paulson School of Engineering and Applied Sciences\\
  Harvard University \\
  29 Oxford St., Cambridge, MA, USA \\
  \texttt{kmahadevan@g.harvard.edu} \\
    \And
 Dr. Michelle C. Yuen \\
  John A. Paulson School of Engineering and Applied Sciences\\
  Harvard University \\
  29 Oxford St., Cambridge, MA, USA \\
    \And
 David T. Farrell \\
  John A. Paulson School of Engineering and Applied Sciences\\
  Harvard University \\
  29 Oxford St., Cambridge, MA, USA \\
    \And
 Prof. Conor J. Walsh \\
  John A. Paulson School of Engineering and Applied Sciences\\
  Harvard University \\
  29 Oxford St., Cambridge, MA, USA \\
    \And
 Prof. Vanessa Sanchez \\
  Department of Mechanical Engineering \\
  Rice University \\
  Houston, TX, USA \\
    \And
 Prof. Robert J. Wood \\
  John A. Paulson School of Engineering and Applied Sciences\\
  Harvard University \\
  29 Oxford St., Cambridge, MA, USA \\
  \And
 Prof. Katia Bertoldi \\
  John A. Paulson School of Engineering and Applied Sciences\\
  Harvard University \\
  29 Oxford St., Cambridge, MA, USA \\
  \texttt{bertoldi@seas.harvard.edu} \\
}

\maketitle

\begin{abstract}

Curved elastic shells can be fabricated through molding or by harnessing residual stresses. These shells often exhibit snap-through behavior and multistability when loaded. We present a unique way of fabricating curved elastic shells that exhibit multistability and snap-through behavior, weft-knitting. The knitting process introduces internal stresses into the textile sheet, which leads to complex 3D curvatures. We explore the relationship between the geometry and the mechanical response, identifying a parameter space where the textiles are multistable. We harness the snapping behavior and shape change through multistability to design soft conductive switches with built-in haptic feedback, and incorporate these textile switches into two wearable devices and one reconfigurable lamp. This work will allow us to harness the nonlinear mechanical behavior of textiles to create functional, soft, and seamless devices. 

\end{abstract}

\vspace{5 mm}
]
\begin{refsection}
\section{Introduction}

Elastic shells -- thin and curved structures \cite{audoly2000elasticity, timoshenko1959theory} -- are ubiquitous in nature and technology, from blood cells and eggs to pressure vessels and roof structures. Across length scales, they are characterized by nonlinear mechanical behavior (e.g., buckling) and are prone to mechanical instabilities \cite{hutchinson2020eml}. While traditionally considered precursors to failure, these instabilities can be harnessed for functionality in elastic shells  \cite{reis2015perspective, koh2023shape}. Leveraging the snapping behavior of elastomeric shells has led to jumpers \cite{gorissen2020inflatable}, swimmers \cite{djellouli2017buckling, lee2022buckling,chen2018harnessing}, programmable fluids \cite{djellouli2024shell}, and passive valves \cite{gomez2017passive}. Elastic shells with a low thickness-to-radius ratio also exhibit two stable configurations, a characteristic that has been exploited to realize switchable optical lenses \cite{holmes2007snapping}, valves for autonomous control of soft actuators  \cite{rothemund2018soft}, and the manipulation of the propagation of transition waves \cite{nadkarni2016unidirectional}.

A variety of strategies are used to fabricate curved shells. Some shells are directly built so that the curved shape is their stress-free state~\cite{lee2016fabrication,jones2021bubble}. In others, curvature arises in initially flat sheets due the residual stresses introduced by growth \cite{pezzulla2016geometry,klein2007shaping}, swelling \cite{pezzulla2015morphing,nardinocchi2015anisotropic}, or mechanical loading \cite{armon2011geometry,chen2012nonlinear}. Complex shapes, including helices and saddles \cite{armon2011geometry, seffen2011prestressed, guo2014shape}, have been realized by adhering elastic sheets stretched in different directions before releasing the resulting multilayer structures. Increasing the complexity of these shapes would require adhering elastic sheets under more complex strain fields, likely requiring artisanal or one-off processes.

A promising method in creating complex surfaces and designer strain fields while maintaining automated and scalable fabrication is to leverage existing textile manufacturing \cite{singal2024programming}. We seek to expand these manufacturing technologies which have been used to create garments for thousands of years to also create wearable devices that make use of multistable curved elastic shells. \newnew{Recent research has begun to shed light on the origins of curvature in knit fabrics, opening opportunities for realization of textiles with complex 3D shapes \cite{niu2025geometric,tajiri2025curling}.} Further, textiles are  flexible, compliant, and can provide protection from the elements \cite{sanchez2021textile}, making them a natural material choice for wearable devices \cite{mahadevan20233d,abel2013hierarchical,o2022unfolding}. Textiles can be used as base materials onto which actuators, sensors, power supplies, an other discrete components are attached, but innovations in structuring including weft-knitting have led to textile-integrated actuators and sensors \cite{sanchez2023knitting,luo2022digital,albaugh2019digital,granberry2019functionally,araromi2020ultra}. These devices are enabled by the fact that the mechanical properties of weft-knits are customizable across the surface; however, it remains an open challenge to formalize rules that translate desired functionality into stitch-level design.
 
In this work, we leverage engineered mechanical instabilities to create functionality in textiles and demonstrate that weft knitting provides a platform to realize multistable textiles. We first introduce pronounced curvature into the knits as a function of stitch pattern. We then explore the relationship between the geometry -- varied periodic stitch patterns -- and mechanical properties -- snap-through instabilities -- of the knits. \newnew{We demonstrate that the curvature of knit textiles arises from pre-stresses introduced during manufacturing, since finite element simulations of bilayers with perpendicular internal stresses reproduce the curved shapes observed in the fabricated knits.} We also investigate the variations in geometry, \new{material, and machine parameters } that result in multistability. Finally, we functionalize these multistable patterns by incorporating conductive elements to create soft switches with built-in haptic feedback. The unique combination of soft, breathable materials and snapping will enable reliable input and interaction with fully textile wearable devices. We envision these switches being integrated into cycling vests for increased visibility and communication while cycling in the dark, without distracting device operation. 
Users can also interact with and reconfigure a responsive soft structure. 

\section{Knit fabrics as perpendicularly pre-stressed bilayers}
\begin{figure*}
  \includegraphics[width=\linewidth]{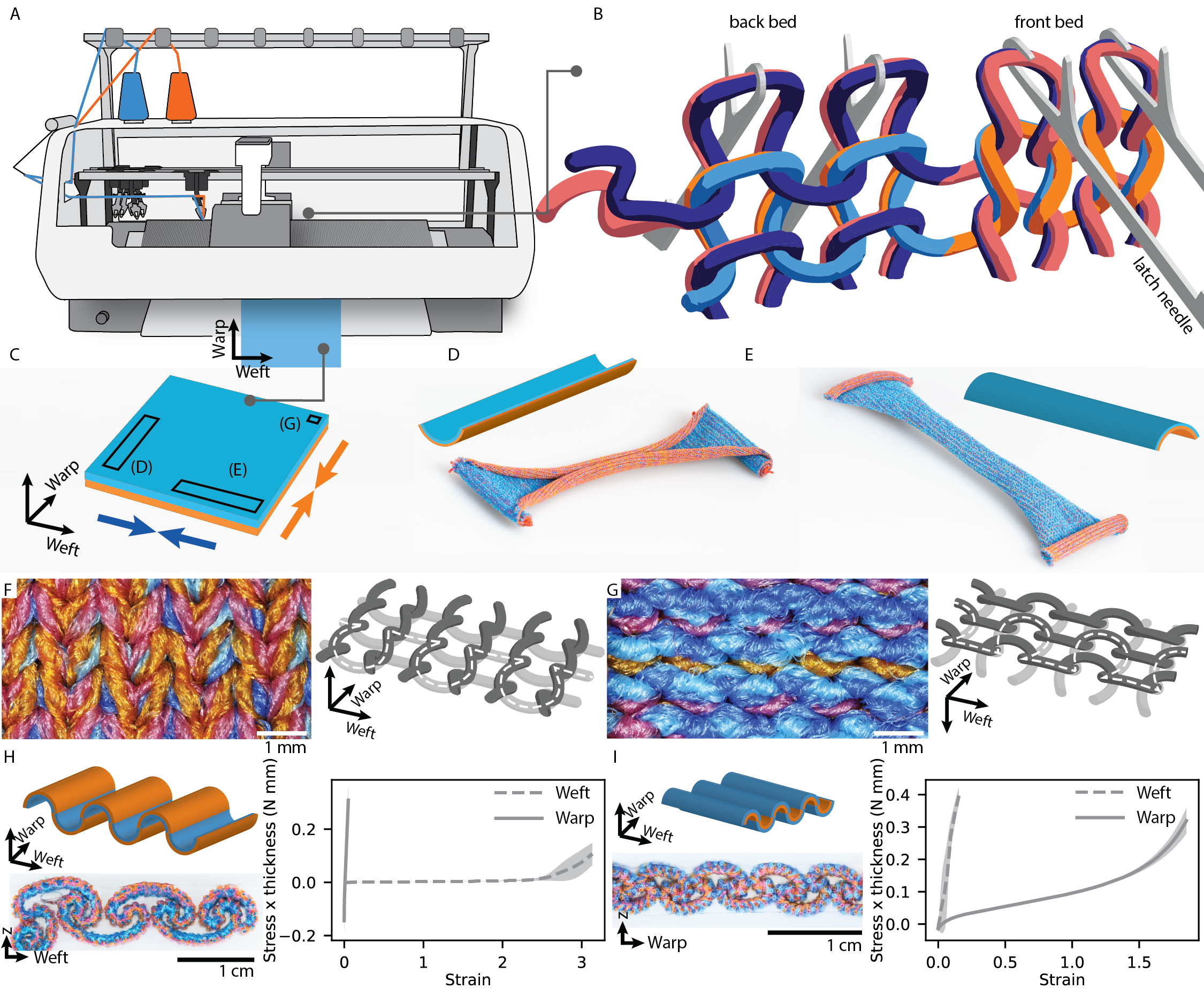}
  \caption{Knitting on the front and back bed of a knitting machine allows us to create corrugated textiles. A) The Kniterate machine. B) Needles are arranged in two parallel beds facing each other. Colored yarns are held together in needles, but offset to color the two faces of the fabric in orange and blue. This needle configuration shows the rib structure in H. C) A knit sample is approximated as a bilayer with perpendicular internal stresses. D) A high aspect ratio sample is cut in the warp direction. A schematic shows the dominant curvature. E) A high aspect ratio sample is cut in the weft direction, resulting in an opposing dominant curvature. F) On the face of the fabric the yarns lie parallel to the warp direction, illustrated with a macro photograph with corresponding schematic. G) On the reverse side the fabric yarns are parallel to the weft direction, shown via imaging and associated schematic.  H) Corrugations in the weft direction are known as a rib fabric. We visualize these corrugations by imaging the cross-section. This fabric is much more compliant in the weft direction as the vertical corrugations unfold. I) Corrugations in the warp direction are known as a garter or links-links fabric. A cross-section image depicts the shape of the corrugations. In a tensile test, these corrugations cause this fabric to be compliant in the warp direction but stiff in the weft direction. \new{All samples are knit on the Kniterate machine.}}
  \label{fig:knit-bilayer}
\end{figure*}
Similar to how a fused filament 3D printer transforms a 1D filament into a 3D structure, a knitting machine structures a 1D yarn into a 2D fabric (Fig. \ref{fig:knit-bilayer}A). This process occurs row by row, through the creation of interlocking slip knots on one of the two mirroring needle beds (Fig. \ref{fig:knit-bilayer}B). All our samples are realized using an automated double-bed weft-knitting machine (Kniterate) and an elastomeric nylon-spandex core-spun yarn (Yeoman Yarns). To visualize complex stitch patterns, we knit with four mechanically identical yarns in different colors (two blue and two orange) such that the face of the resulting fabric is orange and the reverse is blue (Fig. \ref{fig:knit-bilayer}B). All four yarns follow the same looped path, but the orange yarns are offset ahead (a process called plating), resulting in each side of the fabric displaying  a different color. A jersey knit is the simplest of knit structures, where each knit stitch is an identical loop formed on one bed of the knitting machine. 

\newnew{We start by  knitting a jersey fabric and cutting 5 cm $\times$ 20 cm samples aligned with the warp and weft directions (Fig. \ref{fig:knit-bilayer}C). All samples cut from this uniform fabric curl along all four edges, adopting a saddle shape. The sample with its length parallel to the warp direction curls along the long edge with the orange side facing outward (Fig. \ref{fig:knit-bilayer}D), while the sample with length parallel to the weft direction curls along the long edge with the blue side facing out (Fig. \ref{fig:knit-bilayer}E).   The fact that the edges curl when cut demonstrates that internal stresses are present in the knit \cite{armon2011geometry}. The observed  curling  mirrors that of a bilayer elastic sheet with perpendicular internal stresses \cite{armon2011geometry,chen2012nonlinear,pezzulla2015morphing,pezzulla2016geometry}.}
These internal stresses in the fabric result from the manufacturing process. In the knitting machine, tensile loads cause the the yarn to stretch by 105\% (Table \ref{si_fig:yarn_strain}). When released from the needles, the yarn contracts but cannot return to its original state due to the interlocked loop structure, retaining the tensile deformation as internal stress.  \newnew{Examining the yarn paths further supports the analogy between the knits and a bilayer elastic sheet with perpendicular internal stresses; } the yarns on the face of the fabric lie approximately parallel to the warp direction (Figs. \ref{fig:knit-bilayer}F, \ref{si_fig:yarn_paths}), whereas on the reverse side the yarns lie approximately parallel to the weft direction (Figs.~\ref{fig:knit-bilayer}G, \ref{si_fig:yarn_paths}). We can therefore understand the fabric as a bilayer with the technical face and reverse side experiencing perpendicular compressive stresses along the warp and weft directions, respectively. 

Until now, curvature in our samples is concentrated at the sample edges (Figs. \ref{fig:knit-bilayer}D-E, \ref{si_fig:large_edge}). Pronounced curvatures within the bulk are introduced by knitting sections of a textile on the front and back beds of the machine \cite{sanchez2023knitting,abel2013hierarchical} (Fig. \ref{fig:knit-bilayer}B). \new{In particular, we start by knitting  rib and garter (or links-links \cite{rant2014compression,Fashionary2024-ed,Tellier-Loumagne2005-kg}) stitch patterns that consist of warp-direction (vertical) and weft-direction (horizontal) stripes, respectively, where each stripe is a column (rib) or row (garter) alternating between the front and back beds (front-knits and back-knits). \newnew{While the term `links-links' is used to describe the broader category of stitch patterns that contain both front and back bed knits in a single column \cite{rant2014compression,Fashionary2024-ed,Tellier-Loumagne2005-kg}, here we refer to our set of horizontal and vertical striped patterns as garter $n \times n$ and rib $n \times n$, where $n$ is the number of stitches in a front or back bed stripe \cite{sanchez2023knitting}.}
}
 In these striped fabrics, the internal stresses cause corrugations parallel to the stripes (Figs. \ref{si_fig:12 x 12 rib and garter},\ref{si_fig:8 x 8 rib and garter}) which are visible in the cross-section (Figs. \ref{fig:knit-bilayer}H and \ref{fig:knit-bilayer}I). Although both fabrics are corrugated, the shape of the knit stitch reflected in the front-knits and back-knits lead to different distributions of internal stresses that result in different curvature profiles (Fig. \ref{fig:knit-bilayer}H-I). The effect of this pronounced curvature is reflected in the mechanical response of the textiles; the sample is softer when pulled in the direction perpendicular to the stripes since the ridges must first unfold before the textile sheet itself begins to stretch (Fig. \ref{fig:knit-bilayer}H-I,\ref{si_fig:8 x 8 rib and garter}).  In both directions, the response is smooth -- there is no drop in force corresponding to snapping instabilities, -- and reversible -- the sample is monostable and returns to its initial configuration upon unloading. 

\section{Selective patterning can lead to multistability}
\begin{figure*}
  \includegraphics{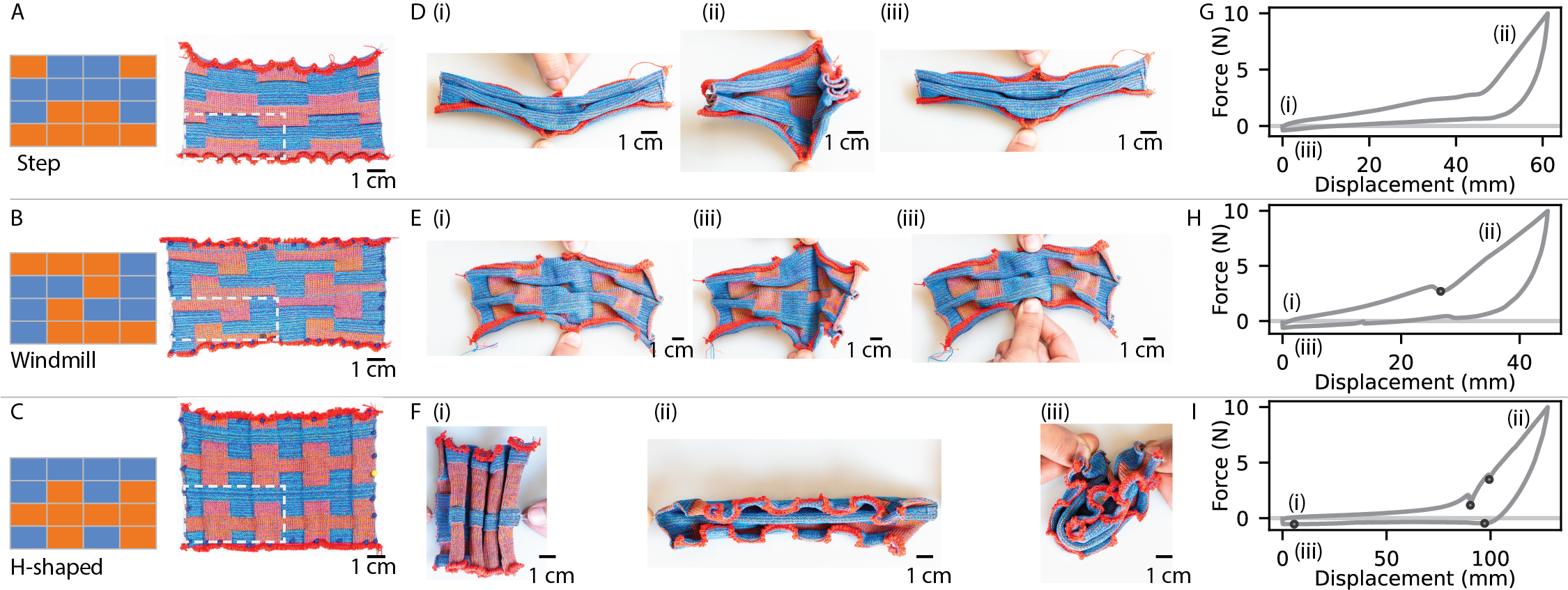}
  \caption{By altering the unit cell knitting patterns we can tune both the final shape and mechanical response of the textile. A) The \emph{Step} pattern unit cell and corresponding knit sample. B) The \emph{Windmill} pattern unit cell and corresponding knit sample. C) The \emph{H-shaped} pattern and corresponding knit sample. D) A \emph{Step} sample (i) before loading, (ii) during loading, and (iii) after unloading. E) A \emph{Windmill} sample (i) before loading, (ii) during loading, and (iii) after unloading. F) An \emph{H-shaped} sample (i) before loading, (ii) during loading, and (iii) after unloading. G-I) Force-displacement measurements for  (G) the \emph{Step} sample, (H) the \emph{Windmill} sample, and (I)  the \emph{H-shaped} sample, with force drops marked and the region of negative force highlighted by a gray horizontal line. \new{All samples are knit on the Kniterate machine.}}
  \label{fig:diff-geometries}
\end{figure*}

\begin{figure}
  \includegraphics{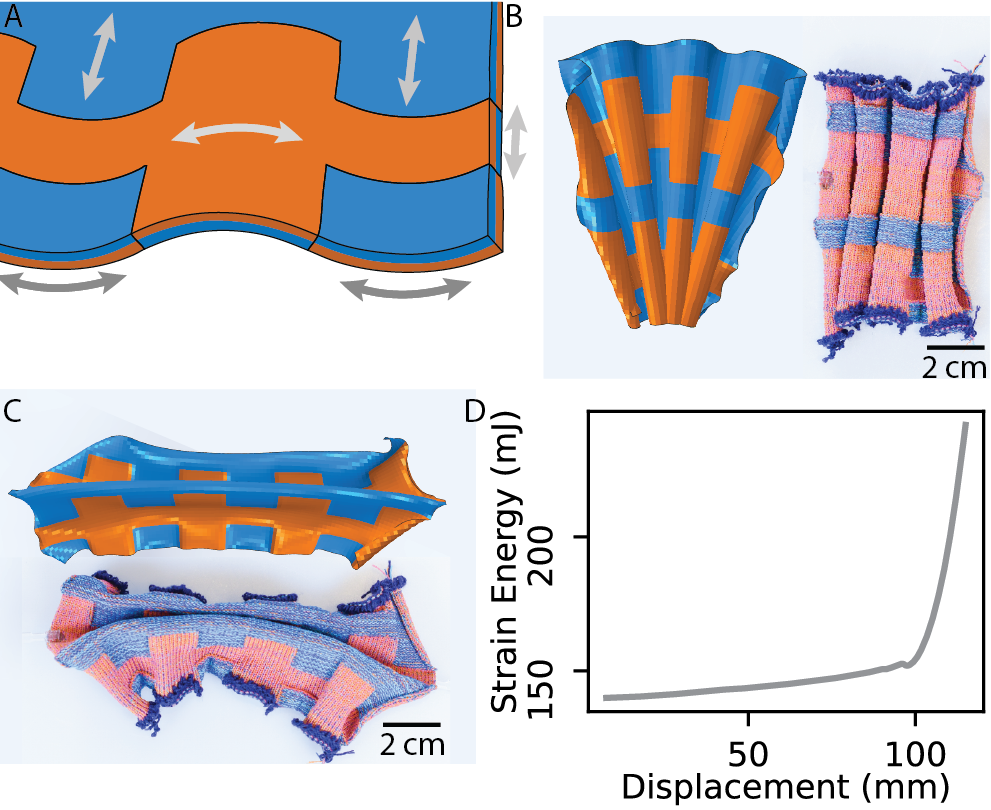}
  \caption{\new{Finite Element simulations demonstrate that the internal stresses in our knits resemble those of a perpendicularly pre-stressed bilayer.
A) Thermal expansion induces curvature and internal stresses in a bi-layer shell with mismatched orthogonal coefficients of thermal expansion.
B) This simplified representation of the fabrics is sufficient to achieve the curved configurations observed in
experiments.
C) When a localized displacement is applied to the boundaries of the model, snapping into a second stable configuration is observed.
D) Numerically predicted evolution of the strain energy as a function of applied displacement exhibits two minima—one at zero displacement and another at a finite, nonzero displacement—confirming multistability.}}
  \label{fig:finite-element}
\end{figure}

We harness internal stresses to generate pronounced local curvatures and exploit them for functionality. By selecting specific knit patterns of front-knits and back-knits, we engineer curved morphologies, snap-through instabilities, and multistability into the fabric. We consider unit cells composed of 4$\times$4 blocks, with each block containing 12$\times$12 stitches knit on either the front or back bed. There are $2^{16}$ possible unit cells that follow these rules, but we focus on three patterns that are balanced, i.e., they have an equal number of stitches knit on the front and back beds: a \emph{Step} pattern  (Fig.~\ref{fig:diff-geometries}A), a \emph{Windmill} pattern (Fig.~\ref{fig:diff-geometries}B), and an \emph{H-shaped} pattern  (Fig.~\ref{fig:diff-geometries}C). By selecting patterns with the same number of stitches on the front bed and back bed, the local curvatures effectively cancel out, resulting in a fabric with no global curvature. We knit samples that are a 2$\times$2 repeat of the unit cells, adding the first row once more at the end as handling regions to facilitate testing.

Upon manufacturing, all fabric samples exhibit pronounced natural curvature in the bulk (Figs.~\ref{fig:diff-geometries}D-i,E-i,F-i). When subjected to point loading, all samples display J-shaped force-displacement curves (Figs.~\ref{fig:diff-geometries}G,H,I, Mov. S1), \new{with an initial low-stiffness region corresponding to unfolding, followed by a stiffening regime associated with stretching \cite{sanchez2023knitting, amanatides2022characterizing}. } Samples also exhibit hysteresis due to internal friction between the yarns. Each pattern exhibits unique mechanical behavior, displayed in the force-displacement curves (Figs. \ref{fig:diff-geometries}G, H, I)  and snapshots taken at key points throughout the loading and unloading curves (Figs. \ref{fig:diff-geometries}D, E, F).

In the \emph{Step} pattern (Fig. \ref{fig:diff-geometries}A), the horizontal garter stripes are perturbed, yet the distinctive horizontal folds remain (Fig. \ref{fig:diff-geometries}D-i). When subjected to a local load by pulling two tabs, the recorded force-displacement curve is smooth, but a vertical fold appears close to the loading point (Fig. \ref{fig:diff-geometries}D-ii), similar to the `scars' seen in corrugated sheets \cite{meeussen2023multistable}. In a uniform corrugated sheet, the fold is determined by only the location and direction of loading, whereas in our sample, the knitting pattern also affects the final configuration. The knit sample is therefore more robust to perturbations in the loading direction. The fabric returns to its initial configuration upon unloading (Fig. \ref{fig:diff-geometries}D-iii).

The \emph{Windmill} pattern disrupts the continuous stripes, and non-perpendicular folds emerge (Fig. \ref{fig:diff-geometries}E-i). Upon loading, the \emph{Windmill} fabric exhibits a sudden drop in force (Fig. \ref{fig:diff-geometries}H), a well-known phenomenon in elasticity  which corresponds to a snapping instability ~\cite{timoshenko1961theory,bazant1991stability}. This snapping event results in the sudden formation of a vertical fold (Fig. \ref{fig:diff-geometries}E-ii). During unloading, we observe a sudden increase in force, corresponding to a snap-back event, where the vertical fold disappears (Fig. \ref{fig:diff-geometries}E-iii) and the fabric returns to its initial configuration.

While the \emph{Step} and \emph{Windmill} patterns are monostable, patterned fabrics can also exhibit multistability \cite{Ahlquist2015}.  For a fabric to be categorized as multistable, there must be a sudden drop in force during loading, which corresponds to a snap-through instability and a transformation into another configuration. 
There must also be no snap-back events during unloading, and a local minimum below zero force, indicating that the second configuration is stable and resists unloading. \new{Both the steep force drop and the below-zero minimum upon unloading are highlighted in Fig. \ref{fig:diff-geometries} with a circular black marker. } An example of such multistable fabrics is the \emph{H-shaped} pattern, which combines the vertical and horizontal stripes of the rib and garter stitches. This sample can fold in the warp, weft, or a combination of both directions, resulting in multiple stable states (Fig. \ref{fig:diff-geometries}F). When the sample is loaded by pulling two tabs, two snapping events are triggered, each corresponding to the disappearance of a warp-direction fold and the appearance of a weft-direction fold (Fig. \ref{fig:diff-geometries}I). The new shape resists unloading, and the force drops below zero during unloading.  While we cannot test all $2^{16}$ patterns, additional examples of both multistable and monostable patterns are shown in Fig \ref{si_fig:checkerboard}. \new{While Fig.~\ref{fig:diff-geometries} presents results for samples subjected to point loads, multistability is also observed under different loading conditions. For instance, when the entire edge of a large \emph{H-shaped} sample is pulled, the structure transitions into a secondary configuration (Fig.~\ref{si_fig:clamped}). However, under these conditions, no snapping behavior is observed in the force--displacement curve, likely due to damping effects along the extended edge.
}
\new{The unique curvature and mechanical response of these knits cannot be predicted from their flat configuration but are instead governed by the pre-stresses introduced during manufacturing. To demonstrate this, we conducted Finite Element simulations using the commercially available software (Abaqus 2021). In these simulations, the knits are modeled as continuum elastic bi-layer shells. Internal stresses originating from the knitting process are replicated in simulation by assigning mismatched orthogonal coefficients of thermal expansion ($\alpha_x=0.165$ $1/C^{\circ}$, $\alpha_y=0.0198$ $1/C^{\circ}$) to the two layers. The simulations consist of two steps; first, the temperature is increased, inducing  internal stresses and curvature. Then we apply a point load to the model. In the layer representing the technical face of the fabric (shown in orange), the thermal expansion causes the layer to expand horizontally, whereas the layer representing the technical back of the fabric (shown in blue) expands vertically (Fig. \ref{fig:finite-element}A). As shown in Fig. \ref{fig:finite-element}B, this simplified representation of the fabrics is sufficient to achieve the curvature shown in experiments. In the second step, we apply a localized displacement to the curved model, mimicking the experimental loading conditions. The simulations reveal snapping and folding transitions corresponding to a secondary minimum in the system’s energy landscape (Fig. \ref{fig:finite-element}C–D), confirming the textile’s intrinsic multistability.}

\begin{figure}
  \includegraphics{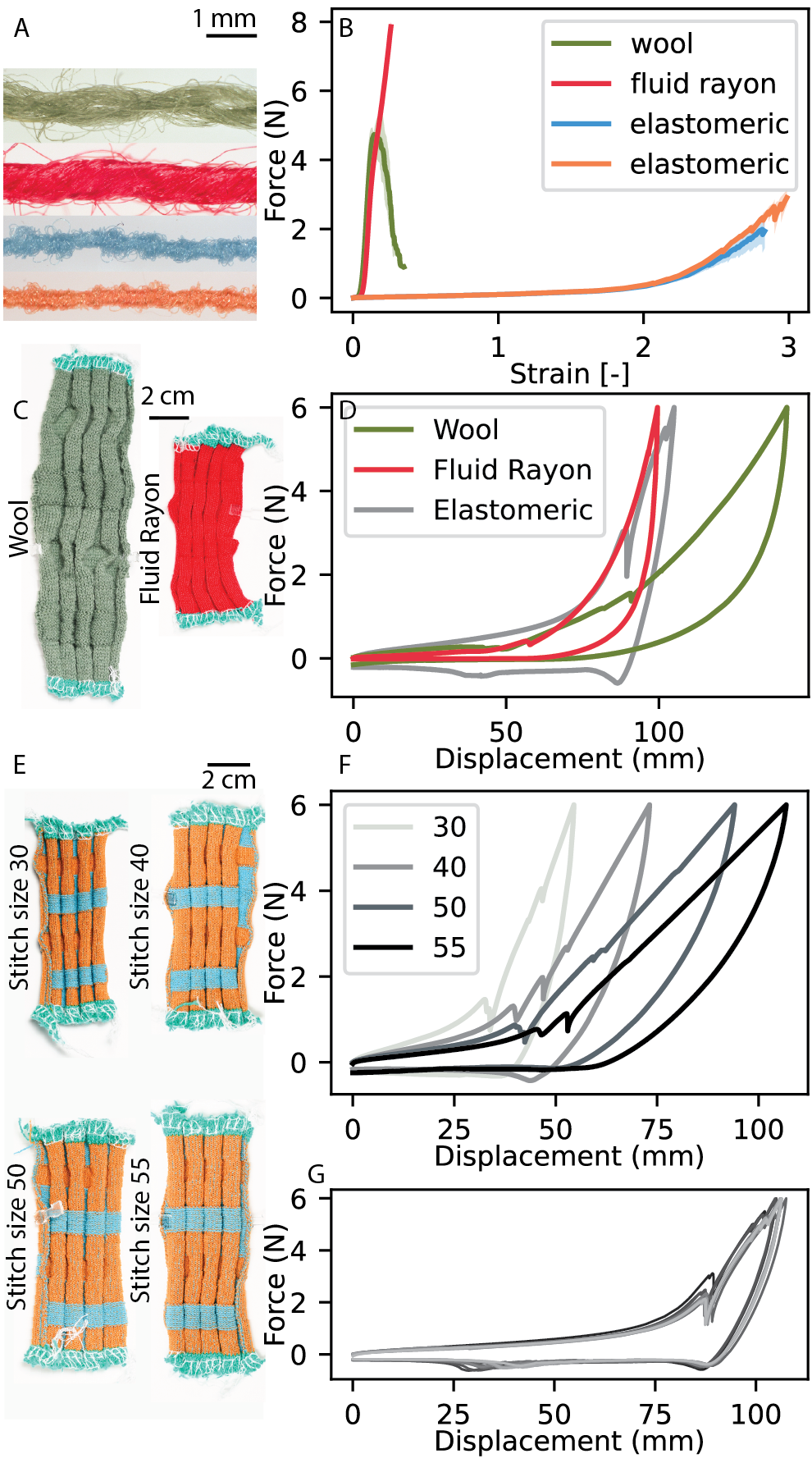}
  \caption{ 
  A) Microscope images of each of the yarns used. B) Tensile tests of the yarns. C) Images of the samples knit from alternate yarns D) Tensile tests of the `H-shape' knit pattern, knit with different yarns. E) Photographs of samples knit with two elastomeric yarns on the Shima Seiki SWG machine in different stitch sizes F) Tensile tests of the `H-shape' knit pattern with different stitch sizes. G) Repeated tensile tests of the sample from Figure \ref{fig:diff-geometries}C, knit with four ends of elastomeric yarn on the Kniterate machine. }
  \label{fig:yarns}
\end{figure}

\section{Tuning multistability}

\new{To investigate how universal this observed multistability is, we knit the \emph{H-shaped} pattern on an alternate automated knitting machine (Shima Seiki SWG N2 15 gauge) and with varied yarns and stitch sizes. We select a 100\% wool yarn (Peter Patchiss) and a yarn called `fluid rayon' that is blend of viscose, spandex, and nylon (Yeoman Yarns), in addition to the elastomeric nylon-spandex yarn we used previously. Each of these yarns are constructed differently, resulting in a different twist and surface texture (Fig. \ref{fig:yarns}A). 
In tensile tests of the yarn alone, we observe that the wool and viscose yarns break at higher forces and are much less extensible than the elastomeric yarns (Fig.~\ref{fig:yarns}B). We knit these yarns using the same knit pattern as Fig.~\ref{fig:diff-geometries}C and compare the mechanical response. However, these yarns are also much thicker and have more loose radial fibers, so we use a larger stitch size to knit them, resulting in differing global shapes (Fig. \ref{fig:yarns}C). Despite the variation in surface, bending, and extensional properties of these yarns, all samples are multistable and exhibit a steep drop in the force-displacement curve during loading, a signature of a snap-through instability (Fig.~\ref{fig:yarns}D). Using two ends of elastomeric yarn (to accommodate the smaller gauge of the machine) and the same knitting pattern in Fig. \ref{fig:diff-geometries}C and Fig.~\ref{fig:yarns}C, we sweep through the range of knittable loop sizes (30-55 in the machine units). As expected, we observe that scaling the stitch sizes leads to scaling the knitted sample (Fig.~\ref{fig:yarns}E). However, despite the fact that the stiffnesses of the samples vary, for all of them we observe the existence of a force drop and, therefore, multistable behavior (Fig.~\ref{fig:yarns}F). Finally, in order to test the repeatability of the response of our knits, we focus on the elastomeric sample tested in Fig. \ref{fig:diff-geometries}C  and repeat the same tensile test 10 times, resetting it manually after each test. As shown in Fig.~\ref{fig:yarns}G, we find that the multistable response of the knits is reversible and repeatable.}

\begin{figure}
  \includegraphics{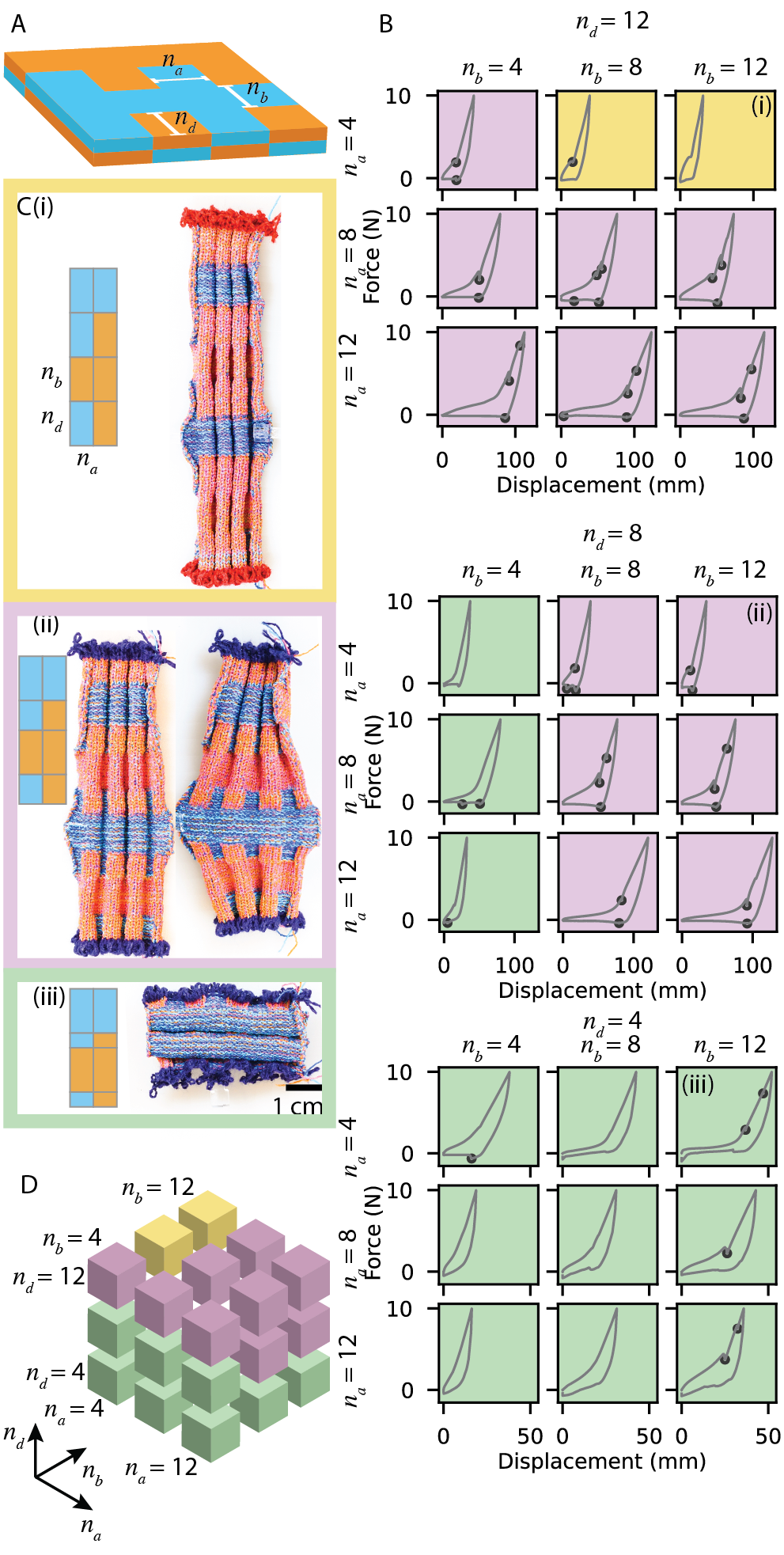}
  \caption{As the geometric parameters of a single unit cell vary, so do the mechanical properties and 3D shapes of the textile. A) Three parameters define our \emph{H-shaped} unit cell ($n_a,n_b,$ and $n_d$). Units of each parameter are number of stitches. B) All three parameters are varied and the force-displacement response measured.  Multistable samples are highlighted in purple, while monostable samples are color-coded based on the dominating folds in their stable configuration: \new{yellow } for the ‘rib’ configuration and \new{green } for the ‘garter’ configuration. C) Snapshots of samples with (i) $n_a$ = 4, $n_b$ = 12, and $n_d$= 12.  (ii) $n_a$ = 4, $n_b$ = 12, and $n_d$= 8. (iii) $n_a$ = 4, $n_b$ = 12, and $n_d$= 4. Additional photos are shown in Figures \ref{si_fig:d12 photos}, \ref{si_fig:d8 photos}, and \ref{si_fig:d4 photos}. D) We visualize which samples are monostable rib -- \new{yellow}, -- monostable garter -- \new{green}, -- and bistable -- purple -- in the 3-D space of our parameters. \new{All samples were knit on the Kniterate machine.} }
  \label{fig:H-shape}
\end{figure}

Next, we control  multistability by tuning the number of stitches in each block that is knit on either the front or back bed. We select the \emph{H-shaped} pattern and vary the block size while maintaining rotational and reflective symmetries. Maintaining symmetries results in three geometric parameters (Fig. \ref{si_fig:hunits}). We vary the number of stitches for the width of the rib, $n_a$, for the height of the garter, $n_b$, and the height of the rib section, $n_d$ (Fig.~\ref{fig:H-shape}A). Setting $n_b$ or $n_d$ to 0 recovers the rib and garter stitch patterns, respectively. We consider twenty-seven unique unit cells with  $n_a$, $n_b$, and $n_d\in [4,\,8,\,12]$. The minimum size of four stitches was selected so that the curvature in the bulk of the samples remains larger than a single stitch. For each geometry, a sample of 2 $\times$ 2 unit cells was manufactured and point-loaded in the same way as shown in Fig.~\ref{fig:diff-geometries}. While geometric parameters can often be reduced to fewer non-dimensional numbers, we demonstrate that all three parameters, $n_a$, $n_b$, and $n_d$, affect the mechanical response separately.

For each of the twenty-seven samples, the force-displacement responses are shown in Figure \ref{fig:H-shape}B. \new{Samples are characterized as monostable or multistable based on the prominence of the force drop during loading and the existence of a force minimum below 0 during unloading. Both of these must be present in order to characterize a sample as multistable. } Multistable samples are highlighted in purple, while monostable samples are color-coded based on the dominating folds in their stable configuration: \new{yellow } for the vertical folds and \new{green } for the horizontal folds (Fig. \ref{fig:H-shape}C). 

In this parameter space of twenty-seven geometries, we observe several trends (Fig.~\ref{fig:H-shape}D). Twelve of the samples we fabricate are monostable in the garter configuration, while two are monostable in the rib. This behavior is because we have selected a unit cell that is a combination of both stripes, but the horizontal garter stripes remain unbroken throughout, whereas the vertical rib stripes are interrupted. We only see the rib folds dominate when they have a small radius of curvature ($n_a = 4$), large area associated with the rib pattern ($n_d = 8$ or 12), and large radius of curvature in the garter direction ($n_b = 12$). 

While it seems possible to reduce these parameters to non-dimensional numbers, the results indicate that multistability requires tuning  $n_a$, $n_b$, and $n_d$ together. For our geometry, setting parameters for the rib sections and garter sections are not equivalent. This could be due to the inherent anisotropic shape of the stitches themselves. The rib sections of the textile are defined by $n_a$, which sets the curvature of the rib folds, and $n_d$, which sets the height of the rib section. The garter section is defined by only one parameter, $n_b$\new{, which sets the height of the horizontal stripe. This stripe height determines the curvature of this garter section}. Therefore the curvature and height of the garter sections cannot be set independently. When $n_b$ is large, the height of the garter section is large, but so is the radius of curvature, leading to less strong folds. As a result, although $n_d/n_b$ defines the ratio between rib and garter stitches, it does not determine whether a fabric is monostable or bistable; samples with the same values of $n_b$ and $n_d$ do not demonstrate the same stability.
\section{Functionality through conductive elements}
\begin{figure*}
  \includegraphics[width=\linewidth]{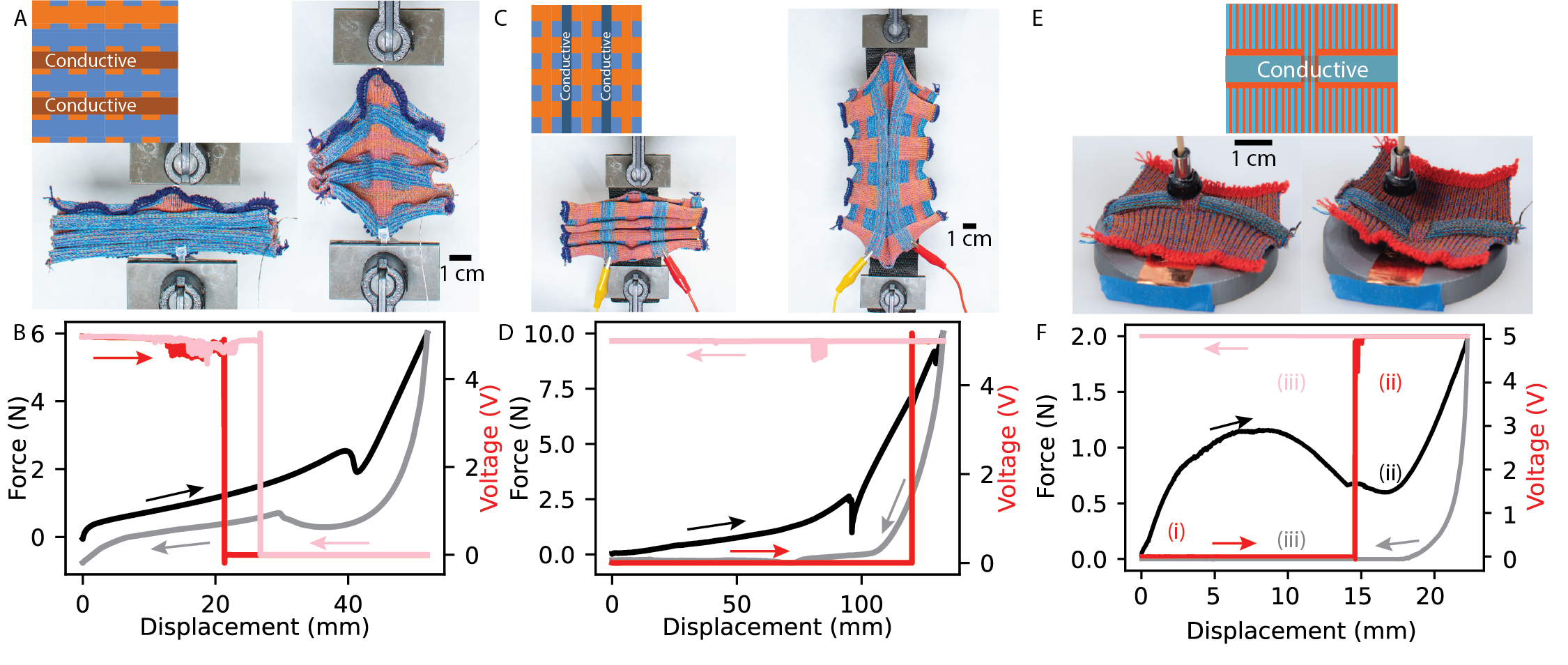}
  \caption{By incorporating conductive ridges, we develop fully textile momentary and latch switches. A) A monostable (momentary) switch consists of an \emph{H-shaped} sample with $n_a = 8$, $n_b = 12$, and $n_d = 4$, and conductive yarn knit into two ridges. Upon stretching, these ridges are separated. B) Voltage and force are both measured as a function of displacement. When the conductive ridges are in contact, the sensor measures 5V. C) A multistable (latching) switch consists of an \emph{H-shaped}  sample with $n_a = 12$, $n_b = 12$, and $n_d = 12$, and conductive yarn knit into two ridges. This sample starts in the rib configuration. When both ridges have snapped into the garter configuration, they are in contact with each other. D) Voltage and force are both measured as a function of displacement. E) An out-of-plane multistable switch is made from the pattern of front and back bed knits that generates a domed button. When the fabric is indented, it snaps into a second stable configuration, and remains there when the load is removed. F) Voltage and force are measured as a function of the indenter displacement. \new{All samples were knit on the Kniterate machine.}}
  \label{fig:switches}
\end{figure*}

We harness both the snap-through behavior and the multistability of the fabric to build a momentary switch and latching switch with built-in haptic feedback. \new{While textile sensors allow for large amounts of scalar or vector data to be collected through unobtrusive means \cite{chaudhari2024characterizing,chen2025self,hang2025multidimensional,cheng2025wearable}, our switches only transmit information about whether they are on or off. Input devices such as a keyboard do not require continuous measurement; a simple binary (on/off) response is beneficial due to less need for data or signal processing. Our devices also provide built-in haptic feedback when they snap between states, giving the user information about their state. } In keyboards, light switches, and controls in cars, haptic feedback allows the user to interact with the built environment without direct visual cues \cite{culbertson2018haptics}. Two key elements of our structure allow us to develop these switches with built-in haptic feedback. First, when a sample transforms from the `rib' configuration to `garter' configuration, the areas of self-contact change. By incorporating a conductive yarn (Fig. \ref{si_fig:conductive yarn}) into the areas that make and break contact, we create a simple switch. Second, this transformation is marked by a steep drop in force which allows a user to know when contact is broken without a visual cue. Humans can sense forces as low as 10-100 mN; the force-drops we demonstrated are on the order of 0.1 N, and therefore detectable \cite{yin2021wearable}.

The momentary switch is realized using an \emph{H-shaped} pattern with $n_a = 8$, $n_b = 12$, and $n_d = 4$, which exhibits snap-through instability but is monostable (Fig.~\ref{fig:H-shape}B). We add conductive yarn to the horizontal ridges that are in contact when the sample is unloaded (Fig. \ref{fig:switches}A). We measure the voltage across this contact as a proxy for electrical conductivity as a function of applied displacement (Fig. \ref{fig:switches}A). We observe that for an applied displacement of $~20$ mm the voltage drops to 0 since the contact between the initially horizontal ridges is broken. After contact is lost, a steep force drop occurs, letting the user know that the switch is on. Since this geometry is not multistable, we observe both a snap-back and the contact resuming upon unloading. A circuit composed of the textile switch in parallel with an LED (Fig. \ref{si_fig:circuit diagrams}) allows us to trigger lighting up of the LED when the contact within the textile is broken (Mov. S2).

A latch switch stays on once activated until it is reconfigured, as do the multistable samples in Fig. \ref{fig:H-shape}. We build a latch switch using an \emph{H-shaped} pattern with $n_a = 12, n_b = 12,$ and $n_d = 12$ that is multistable. The sample displays two force-drops which correspond to the formation of two horizontal ridges (Fig. \ref{fig:switches}C). These two ridges have conductive yarn, and when they come into contact the voltage jumps to 5V. This contact occurs at a displacement between the two force drops; after the second force drop, the user knows the light is on (Fig. \ref{fig:switches}D).

Until now, we focus on periodic geometries that snap under tension. However, our strategy can be generalized to non-periodic geometries. We demonstrate an example that behaves like a button, i.e., it snaps under a local compressive load and comes into contact with a plate below it. The button is designed to rise above a flat stitch pattern (1$\times$1 rib - Fig.~\ref{fig:switches}E). A single stripe of garter interrupts this flat stitch, forming a ridge. Continuing the rib pattern through the center of the ridge transforms it into a button (Fig.~\ref{fig:switches}E-i). When pushed, this dome switches to its inverted stable configuration and comes into contact with copper tape to complete the circuit (Figs.\ref{fig:switches}E-ii,E-iii). This latch switch goes from ``off'' to ``on'' at the lowest point in the force-displacement curve, unlike the other two examples, and therefore the user can detect exactly when the switch turns on (Fig.~\ref{fig:switches}F, Mov. S2).

\begin{figure*}
  \includegraphics{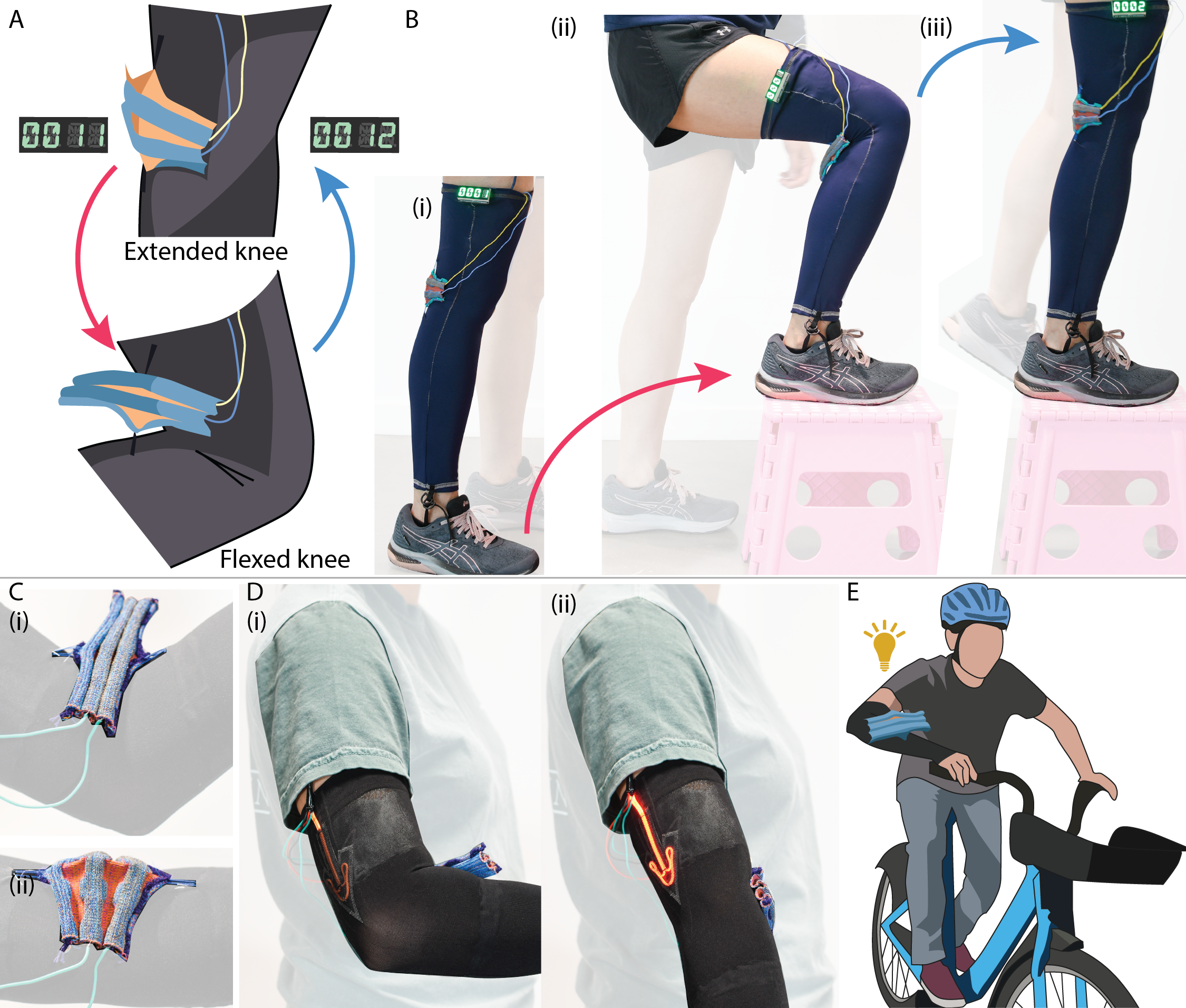}
  \caption{The momentary switch is further developed into a step counter and turn signaling sleeve for cycling. A) The knit counter is attached to a sleeve by threads such that the opening and closing of the knee leads to the knit counter being point loaded. An attached Arduino microcontroller is able to count one full cycle of the counter opening and closing and display the total number of steps. B) The whole device is shown to count one step when a user steps onto a stool. C) The same knit textile is attached to the elbow joint, and is in the closed configuration when the elbow is flexed (i) and the open configuration when the elbow is extended (ii). D) An attached circuit causes an LED arrow to switch on when the user signals to turn. E) This device can be used to signal turns while cycling, allowing for increased visibility and safety. \new{All samples were knit on the Kniterate machine.}}
  \label{fig:step-counter}
\end{figure*}

\begin{figure}
  \includegraphics[width=\linewidth]{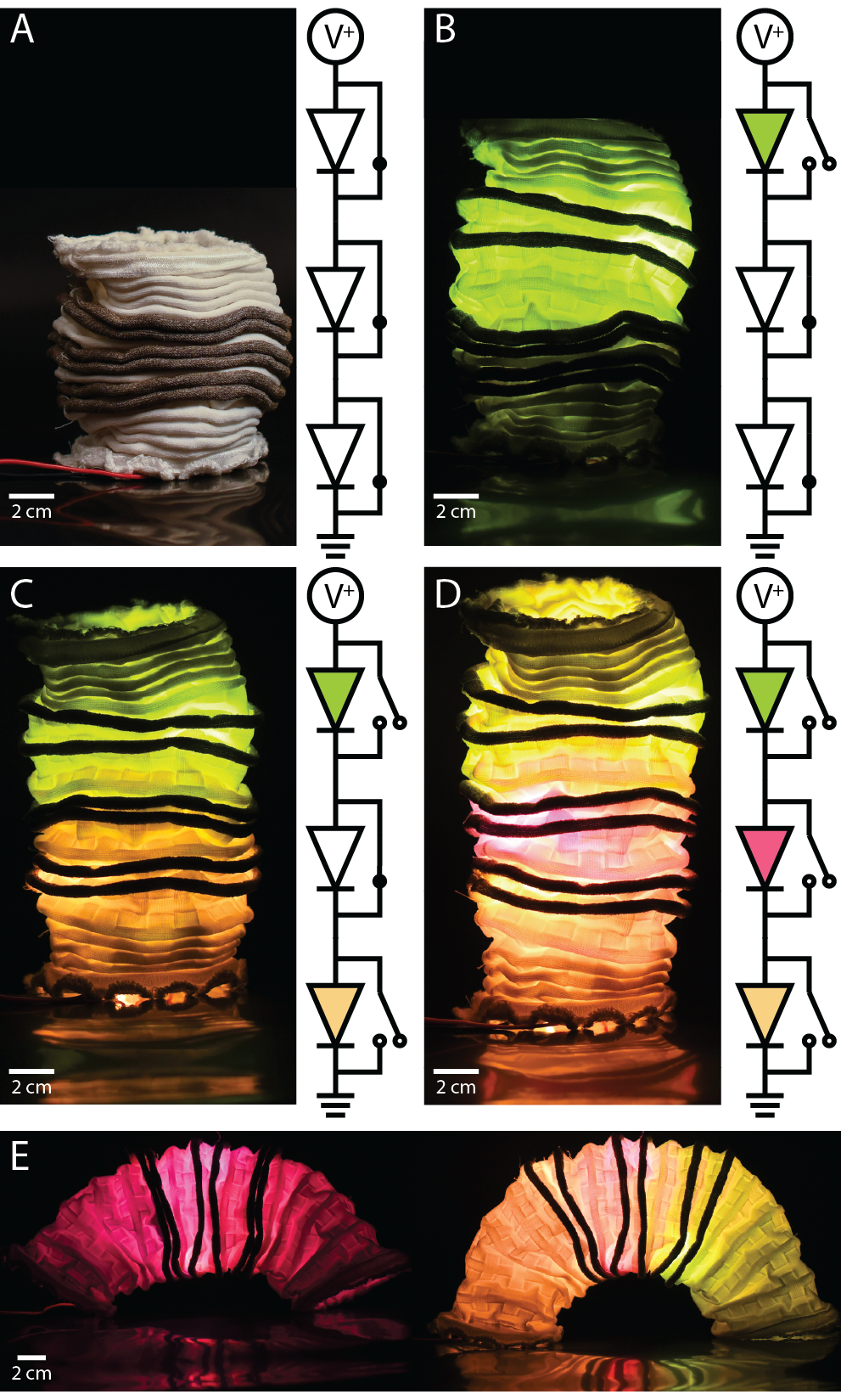}
  \caption{We combine three latch switches to create a reconfigurable lamp. A) The lamp is off. All three switches are closed. B) One, C) two, and D) three switches are open, corresponding to one, two, and three LEDs switching on. E) The multistability of the structure allows a user to reconfigure the switches and the global shape. \new{All samples were knit on the Shima Seiki SWG machine.}}
  \label{fig:lamp}
\end{figure}
A natural application for these textile switches is in wearable devices. By placing a momentary switch across the inside of a user's knee (Fig. \ref{fig:step-counter}A), we can detect whether the knee is flexed (bent) or extended (straight). When the knee is flexed, the two conductive ridges come into contact with each other, and a circuit is closed. When the knee is extended, this circuit remains open. An Arduino microcontroller records each cycle of closing and opening as a step, and the total number of steps is displayed on the built in display (Fig. \ref{fig:step-counter}B). The knit piece here acts as a binary sensor which simplifies the complex motion involved in a step into a simple cycle of ``on'' and ``off'' (Mov. S3). 

In addition to this lower limb device, we also place the momentary switch across the elbow joint in order to detect whether the elbow is flexed or extended. The ridges are in contact while the user's elbow is bent and contact is broken when the user's arm is straight (Fig. \ref{fig:step-counter}C). The switch is incorporated into a sleeve along with a series of lights that display an arrow when the arm is straight (Mov. S3). This allows the user to signal a turn while increasing visibility during cycling at night. \new{The user can communicate using the existing hand signal for turning while cycling, and } the snapping behavior of the switch allows the user to feel whether the light is on or off without visual feedback or confirmation. Analogous to haptic feedback in car buttons and switches, this allows the cyclist to focus their eyes and ears on the road since the device inputs through the sense of touch. The whole device remains soft and breathable, important characteristics for a garment that will be worn during high output exercise. \new{This device provides haptic feedback while allowing a cyclist to use established hand signals.} 

While popular in wearables, textiles are also found in household items and furnishings. They are easy to clean, comfortable, and diffuse light in lamps and window coverings. With this in mind, we develop an adaptable textile lamp in which multistability enables shape reconfiguration and serves as an electronic latch switch. To form this lamp, we sew a rectangular panel of three parallel latch switches into a cylinder. \new{These panels were all knit on the Shima Seiki SWG machine for larger size and increased reliability. } Each switch consists of two rings that come into contact with each other. An LED is placed in parallel with each switch so that the light shines when the switch is open. In Figure \ref{fig:lamp}A, all three switches are closed, and the lights are off. Any of the eight combinations of three LEDs are possible -- we demonstrate three in Figures \ref{fig:lamp}B-D. The entire textile is multistable, and the switches wrap around the whole cylinder, allowing this lamp to be reconfigured into any shape and combination of ``on'' lights (Fig.~\ref{fig:lamp}E, Mov. S4). As the structure changes shape, individual unit cells snap back and forth, providing haptic feedback that help the user to understand how the lighting will respond. This textile lamp is also lightweight and portable, weighing 71.4g including all electronics and fits into a single hand. Designed and knit on an industrial machine, the textile portion of this lamp takes 25 minutes to knit and is scalable, paving the way for interactive, morphing textiles for the home.

\section{Conclusion}

To summarize, we have \new{presented a perspective on  modeling the curvature of a knit structure as a perpendicularly pre-stressed bilayer. } This approach allows for the creation of complex curved knit surfaces that exhibit snap-through instabilities and multistability. \newnew{In the future, a more detailed investigation into the interactions and stresses in the individual yarns would allow for the creation of an analytical continuum model of these textiles.} We demonstrated that the onset of multistability can be tuned by altering both the shape and dimensions of the knit patterns, \new{and that this multistability is robust across materials and knitting parameters}. Furthermore, functionality is added by incorporating conductive areas, enabling the design of momentary and latch switches with built-in haptic feedback. These switches open avenues for soft wearable devices that users can interact with without the need for external motors or visual feedback. With the advent of `the internet of things,' engineers and artists are searching for new ways to interact with the natural and built worlds, and these lightweight, breathable, and compliant switches can be integrated into garments and wearable robots without additional bulk or weight. We have demonstrated a switch across both the elbow and knee joints. These switches can be adapted to be triggered in response to an uneven gait or terrain to provide feedback to controllers in assistive devices that provide additional support or stabilization. Wearables are not the only textiles we interact with on a daily basis; we also demonstrate an interactive textile lamp which can be reconfigured to produce difference lighting effects and shapes. These haptic interfaces can also be incorporated into upholstered interiors, including cars and furniture. 
\new{While this study explores only a small portion of the vast design space of knit patterns, future work can build on this framework by extending our Finite Element model to predict the \newnew{exact } curvature and mechanical properties of a broader range of knit patterns, enabling the creation of a new class of functional and wearable devices.} 

\vspace{5 mm}
\textbf{Methods}\par
All knit samples were fabricated on either a Kniterate knitting machine or a Shima Seiki SWG N2 061 using knitout, a Python frontend, and Kniterate backend \cite{mccann2016compiler,KnitoutFrontendPy,KniterateBackend}. The details of the materials and fabrication methods are further summarized
in Supplementary Information sections 1 and 2. The experimental
procedure of the tensile testing of textile samples are in Supplementary Information section 3. The details of the Finite Element simulations are in the GitHub repository For reference, all relevant codes are available in the accompanying GitHub repository (\href{https://github.com/bertoldi-collab/multistableknits_FE}{link}). Finally, additional data showing repeated experiments and additional knit patterns are shown in Supplementary Information section 4.
\medskip

\medskip
\textbf{Acknowledgements} \par 
Research was supported by the NSF grant DMR-2011754 and ARO MURI program W911NF-22-1-0219.  Equipment was supported by ONR DURIP Award N00014-19-1-2220. K.B. also acknowledges support from the Simons Collaboration on Extreme Wave Phenomena Based on Symmetries. V.S. also acknowledges salary support from NSF grant DMR-2138020 for the MPS-Ascend program. The authors would like to thank the Carnegie Mellon Textiles Lab and Prof. Jim McCann for discussions and resources relating to machine knitting. They also thank Dr. Anne Meeussen, Helen E. Read, Prof. Ahmad Rafsanjani, and Prof. Antonio E. Forte for insightful conversations and feedback. 

\medskip

\printbibliography[title={References}]
\end{refsection}
\newpage
\onecolumn



\smallskip

\setcounter{section}{0}
\setcounter{figure}{0}
\setcounter{equation}{0}
\renewcommand{\thefigure}{S\arabic{figure}}
\renewcommand{\thetable}{S\arabic{table}}
\renewcommand{\theequation}{S\arabic{equation}}
\renewcommand{\thesection}{S\arabic{section}}
\begin{refsection}

\section{Materials}

\textit{Elastomeric Yarns:} Elastomeric yarn samples were produced using four ends of Yeoman Yarns Elastomeric Nylon Lycra (Yeoman Yarns, United Kingdom), made from 81\% nylon and 19\% Lycra (a brand name of spandex), and plated together (i.e., knit together as one yarn). 

\textit{Conductive Yarn:} Two ends of conductive yarn were added to the existing four ends of elastomeric yarn in order to create conductive sections of textile. The conductive yarn used is `20 dtex Lycra + 44Ag + 44 roh Twisted Yarn', purchased from V Technical Textiles, Palmyra, NY.  \\

\noindent\textit{Acrylic Waste Yarn:} Three rows of acrylic yarn (16/2 Vybralite Acrylic Yarns, National Spinning Co. (Peter Patchis Yarns, USA)) were knit to start and end each sample to facilitate fabrication and prevent unraveling.  \\
\begin{figure}[h]
\centering
  \includegraphics{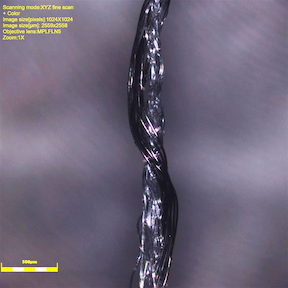}
  \caption{A microscope image shows the conductive fibers twisted around a Lycra core. Scale bar is 500 $\mu$m.}
  \label{si_fig:conductive yarn}
\end{figure}
\section{Fabrication}

\subsection*{Automated knitting}
All samples were knit on the Kniterate knitting machine with stitch size 5 and four ends of elastomeric yarn fed as one. Conductive areas of textile were fabricated by threading a separate feeder with four ends of elastomeric yarn and two ends of conductive yarn. This feeder was used to knit the rows required to be conductive. 

Through a series of springs, the yarn fed into the Kniterate mach
ine is held under tension. We approximate the strain applied by the machine as follows: While under tension in the machine, five sections of each yarn 10 cm in length are measured and  cut, which releases the tension \cite{ohlson2023estimating}. The relaxed length of these yarn sections is then measured with a ruler and the tensile strain calculated by 
\begin{equation}
    \varepsilon = \frac{10-l_0}{l_0},
    \label{si_eq:yarn strain}
\end{equation}
where $l_0$ is the measured length (in centimeters) of each yarn section once the tension is removed. The tensile strains are reported in Table \ref{si_fig:yarn_strain}. The average strain across all colors and sections is 1.05.

\begin{table}[]
\centering
\caption{Strain approximation of the yarn in the machine. Five 10 cm linear tracts of each yarn are cut while under tension in the machine. The relaxed length of these yarn sections is then measured with a ruler and the tensile strain calculated as in Equation \ref{si_eq:yarn strain}. The average across all colors and sections is 1.05}
\begin{tabular}{l|r|r|r|r}
                                                                                                                        & \multicolumn{1}{l|}{\textbf{Orange 1}} & \multicolumn{1}{l|}{\textbf{Orange 2}} & \multicolumn{1}{l|}{\textbf{Blue 1}} & \multicolumn{1}{l}{\textbf{Blue 2}} \\ \hline
\multirow{5}{*}{\begin{tabular}[c]{@{}l@{}}Tensile strain\\ imposed by\\ knitting machine\\ {[}mm / mm{]}\end{tabular}} & 1.63                                   & 1.17                                   & 0.89                                 & 0.79                                \\
                                                                                                                        & 1.56                                   & 1.04                                   & 0.85                                 & 0.82                                \\
                                                                                                                        & 1.56                                   & 1.08                                   & 0.82                                 & 0.85                                \\
                                                                                                                        & 1.50                                   & 1.08                                   & 0.82                                 & 0.82                                \\
                                                                                                                        & 1.33                                   & 0.92                                   & 0.67                                 & 0.89                                \\ \hline
Mean                                                                                                                 & 1.52                                   & 1.06                                   & 0.81                                 & 0.83                                \\ \hline
Std Dev                                                                                                                 & 0.12                                   & 0.09                                   & 0.08                                 & 0.04                               
\end{tabular}
\label{si_fig:yarn_strain}
\end{table}
Each stitch pattern shown was defined by a grid of black and white squares in Adobe Photoshop. These images were then read into python, where they were converted into knitout code using the python frontend \cite{KnitoutFrontendPy}. The resulting knitout files were then compiled to `.kc' files through the knitout-to-kniterate backend \cite{KniterateBackend}, which are the format required by the kniterate machine. Three rows of acrylic waste yarn were knit onto the start and end of each sample for ease of knitting as well as for handling purposes (e.g., to prevent unraveling of the samples' testing region). 

\subsection*{Device fabrication}
\begin{figure}
\centering
  \includegraphics[width=\linewidth]{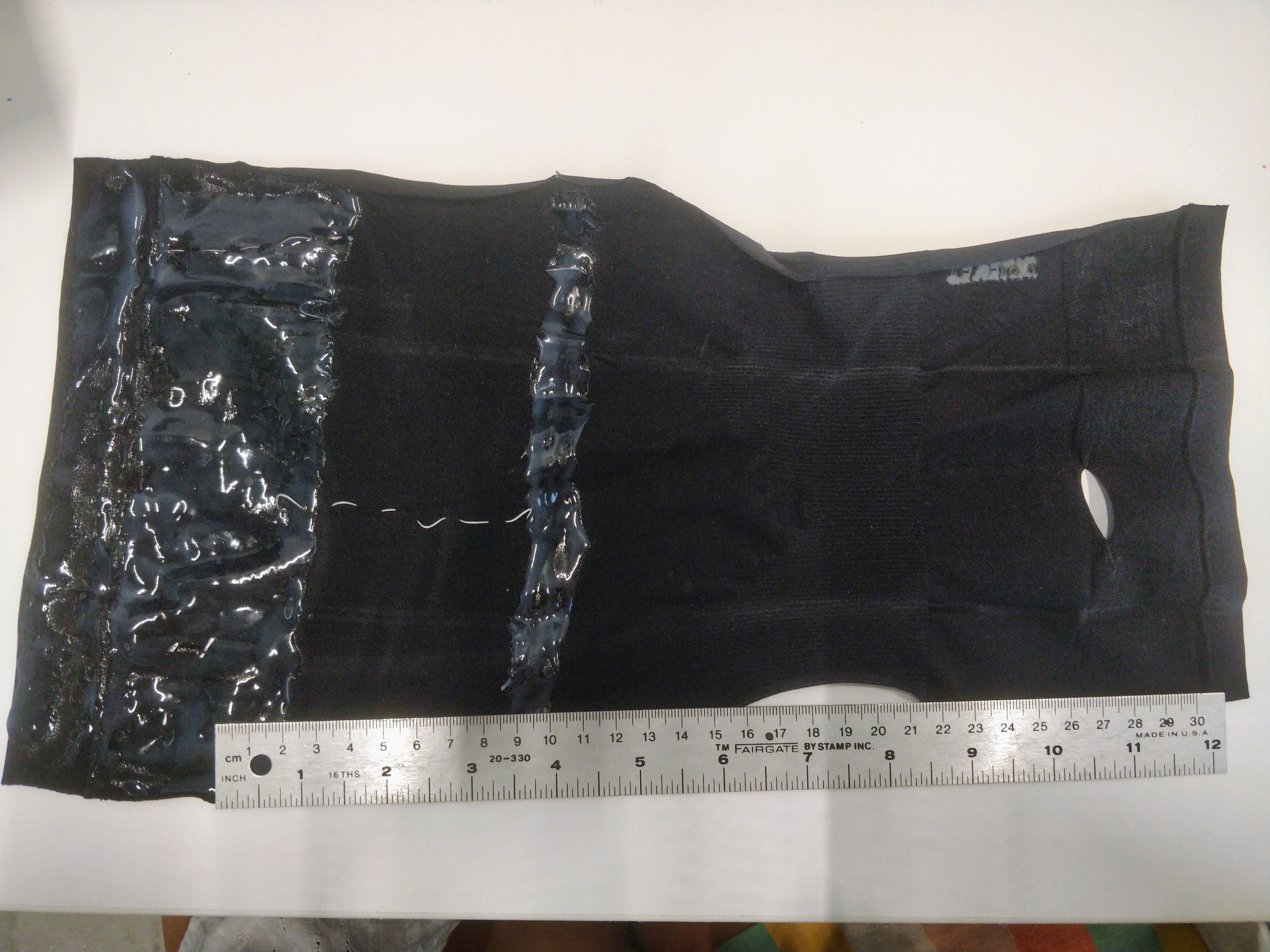}
  \caption{The inside of the turn-signal sleeve during the fabrication process shows the silicone (DragonSkin 10, Smooth-On) coating before addition of circuitry and re-assembly.}
  \label{si_fig:sleeve}
\end{figure}

\textbf{Turn-signal wearable device.} The turn-signal wearable device described in Fig. \ref{fig:step-counter} of the main text was fabricated onto a base of a cooling compression sleeve with a thumbhole made from 4-way stretch fabric (Teveuni, China). The sleeve was cut open to assemble the device. Dragonskin 10 was applied in a thin layer on the inside of the sleeve on either side of the elbow joint to anchor the sleeve to the wearer and provide better force transfer between the knit sample and the wearer (Fig. \ref{si_fig:sleeve}). A flexible LED filament (Adafruit) was connected via wire and conductive thread to the textile switch, which was hand-sewn across the elbow joint. After the circuit was assembled, the sleeve was stitched back into a tube on a five-thread overlock machine. \new{This device can be powered by either a desktop power supply or portable USB-C power bank, which can weigh as little as 250g.}

\textbf{Lamp.} The knit portion of the lamp described in Fig.~\ref{fig:lamp} of the main text was fabricated by knitting a bistable substrate with three paired bands of conductive textile, where each serves as one of three switches. These three switches were each individually paired with a commercially available flexible LED filament in parallel, and then each switch-LED pair was connected in series. When each switch was closed, its corresponding LED filament would turn off as the current would pass through the switch; when each switch opened, its LED filament would illuminate. The circuit was powered with a DC power supply set at 7.5V and current-limited at 0.04A.

\textbf{Step counter.} The step counter device described in Fig.~\ref{fig:step-counter} was fabricated using a base of a Spandex fabric tube, which was then instrumented with a monostable knit sample attached spanning the inside of the knee joint. The two conductive regions of the knit sample form a switch that would repeatedly make and break contact with each step. This switch state was monitored with a digital input pin on an Arduino Nano Every, with the step count being incremented each time the switch closed during knee bending. \new{This device is powered by a portable USB-C power bank, which can weigh as little as 250g and fits into a pocket.}

\section{Testing}

Rib, garter, and jersey samples were tested on an Instron tensile testing machine with boundaries fixed by a set of hook grips previously described \cite{sanchez2023knitting}. Samples with complex curvature are point-loaded in order to isolate and  observe local deformation defining multistability. Point loading is accomplished by gluing a 5mm x 15 mm acrylic tab to each sample using Loctite 416 instant adhesive. The stiff acrylic tabs allow force to be transferred from the textile sample back to the load cell, allowing us to measure negative forces.

\begin{figure}
  \centering
\includegraphics{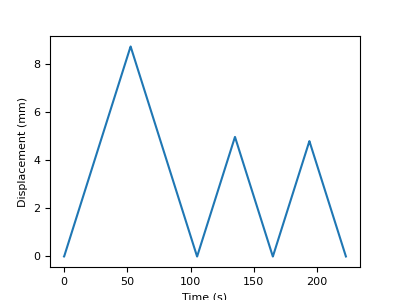}
  \caption{An example of the displacement-controlled protocol for tensile testing of rib, garter, and jersey samples. The first cycle's peak is at 25 N, and the second and third cycle's peaks are at 15N. The second cycle of multiple samples is averaged and presented in Figures \ref{fig:diff-geometries}, \ref{si_fig:12 x 12 rib and garter}, \ref{si_fig:8 x 8 rib and garter}, \ref{si_fig:jersey}}.
  \label{si_fig:time-displacement}
\end{figure}

All samples are loaded and unloaded at 10 mm/min. The rib, garter and jersey samples are cycled three times, up to 25N, 15N , and 15 N (Fig. \ref{si_fig:time-displacement}). The samples with complex geometry (i.e., \textit{Step}, \textit{Windmill}, \textit{H-shaped}) are loaded through a single cycle up to 10N. 

\begin{figure}
  \includegraphics[width=\linewidth]{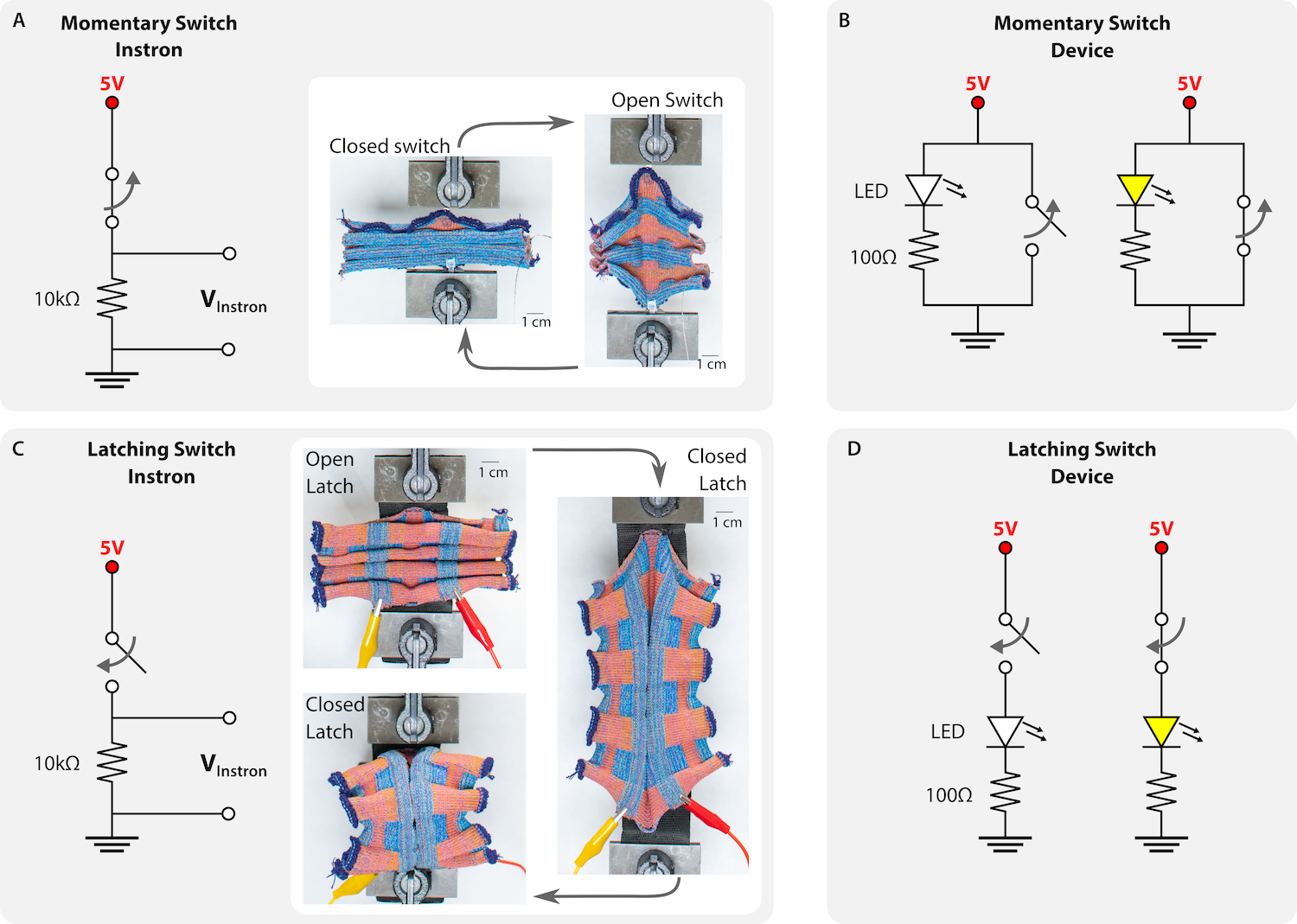}
  \caption{Circuit diagrams for the momentary (A,B) and latching (C,D) switches for measuring change in voltage in mechanical tests (A,C), and when integrated into wearable garments (B,D).}
  \label{si_fig:circuit diagrams}
\end{figure}

For samples where voltage is measured, a circuit connects the sample to a power source and the Instron tensile tester, enabling synchronized recording of  voltage, force, and displacement (Fig. \ref{si_fig:circuit diagrams}).

\clearpage
\newpage
\section{Finite Element simulations}
To demonstrate that the  unique curvature and mechanical response of the considered  knits  are  determined by the pre-stresses introduced during manufacturing, we performed Finite Element (FE) simulations using the commercially available software Abaqus 2021 \cite{abaqus2021}. Although developing numerical tools capable of accurately capturing the behavior of textiles remains an ongoing challenge, we show here that a simplified bilayer shell model can effectively elucidate the fundamental mechanisms driving the mechanical response of the knits under consideration. 

\begin{figure}
\centering
  \includegraphics{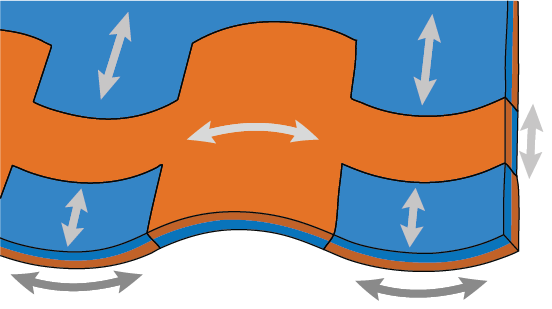}
  \caption{Directions of expansion in our bilayer simulation. The orange layer (technical face) expands in the x-direction, whereas the blue layer (technical back) expands in the y direction.}
  \label{si_fig:expansion}
\end{figure}

All models are composed of an initially flat rectangular slab discretized with 4-node quadrilateral shell elements featuring hourglass control and reduced integration (S4R elements). Each model includes a composite bilayer section. The internal stresses arising from the knitting process are replicated in the simulation by assigning different coefficients of thermal expansion to the two layers. Specifically, both layers share identical linear elastic material properties but differ in their thermal expansion coefficients: ($\alpha_x$, $\alpha_y$) = (0.165 $1/°C$, 0) for the orange layer (technical face) and ($\alpha_x$, $\alpha_y$) = (0, 0.0198 $1/°C$) for the blue layer (technical back) in Fig. \ref{si_fig:expansion}.

Due to the instabilities triggered during the simulations, the finite-sized samples were analyzed using a nonlinear dynamic implicit procedure (*DYNAMIC) under quasi-static conditions. Rayleigh damping was applied with proportional mass damping defined by a coefficient of $\alpha_r = 4$ for the thermal expansion step. For the subsequent point loading step this damping was scaled based on the approximate experimental time ($\alpha_r = 1400$). Critically, it is noted that the multistability of all samples is highly dependent on this damping parameter and can be tuned to occur even in the Step sample by reducing this damping value. No nodal imperfections were introduced, as the natural asymmetric pre-stress across the shell section was sufficient to initiate the deformation. For reference, all relevant codes are available in the accompanying GitHub repository (\href{https://github.com/bertoldi-collab/multistableknits_FE}{link}). Each simulation consisted of two steps:

\subsection*{Step 1: Thermal Expansion}
In the first step, the temperature is increased by one degree, generating internal stresses and curvature. The temperature change serves to replicate the internal stresses introduced during manufacturing. To ensure uniform thermal contraction, periodic boundary conditions were applied along the cell boundaries \cite{Danielsson.2002, Farrell_2025}. 


\begin{figure}[h]
\centering
  \includegraphics{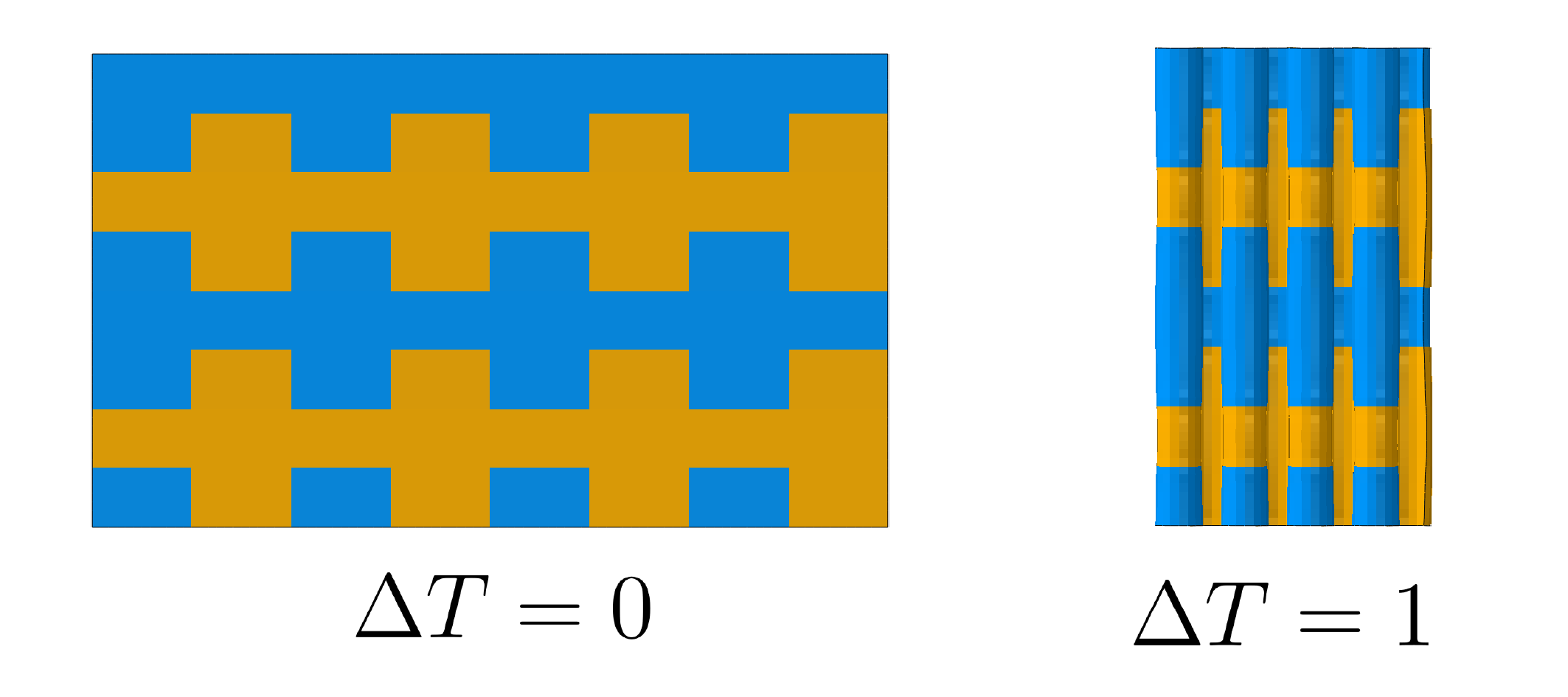}
  \caption{FEM contraction and curving of the knit simulated H-shaped pattern in response to thermal loads. }
  \label{si_fig:sim_boundaries}
\end{figure}


When subjected to thermal loading, the shell models buckle out of plane, forming curved profiles as illustrated in Fig. \ref{si_fig:sim_boundaries}. The resulting curvature of the simulated models depends primarily on the thermal expansion coefficients and the thickness of the shell section. In this study, the thermal expansion coefficients of the two layers ($\alpha_x$, $\alpha_y$) were adjusted to match the curvature observed in the manufactured knitted samples. The specific thermal expansion values used for each sample are listed in Table \ref{tab:thermal_expansion}.

\begin{table}[h!]
\centering
\caption{Orthogonal thermal expansion coefficients for each sample geometry.}
\begin{tabular}{lcc}
\hline
\textbf{Sample Type} & Orange layer  $\boldsymbol{[\alpha_x},\boldsymbol{\alpha_y]}$ (C$^{-1}$) & Blue layer $\boldsymbol{[\alpha_x},\boldsymbol{\alpha_y]}$ (C$^{-1}$) \\
\hline
H-shaped   & [ 0.165, 0 ] & [ 0, 0.0198 ] \\
Windmill   & [ 0.04,  0 ] & [ 0,   0.165 ] \\
Step       & [ 0.04,  0 ] & [ 0,  0 .105 ] \\
\hline
\end{tabular}
\label{tab:thermal_expansion}
\end{table}

\subsection*{Step 2: Mechanical Loading}
In the second step, a mechanical load is applied to the model. To replicate the boundary conditions of the finite-sized samples shown in Fig. \ref{fig:diff-geometries}, the periodic boundary conditions are first removed, and the model is allowed to relax. After relaxation, a displacement is gradually applied to a few boundary nodes (point loading) to reproduce the experimental conditions depicted in Fig. \ref{fig:diff-geometries}. 

\subsection*{Numerical results}
\begin{figure}[h]
\centering
  \includegraphics[width=\linewidth]{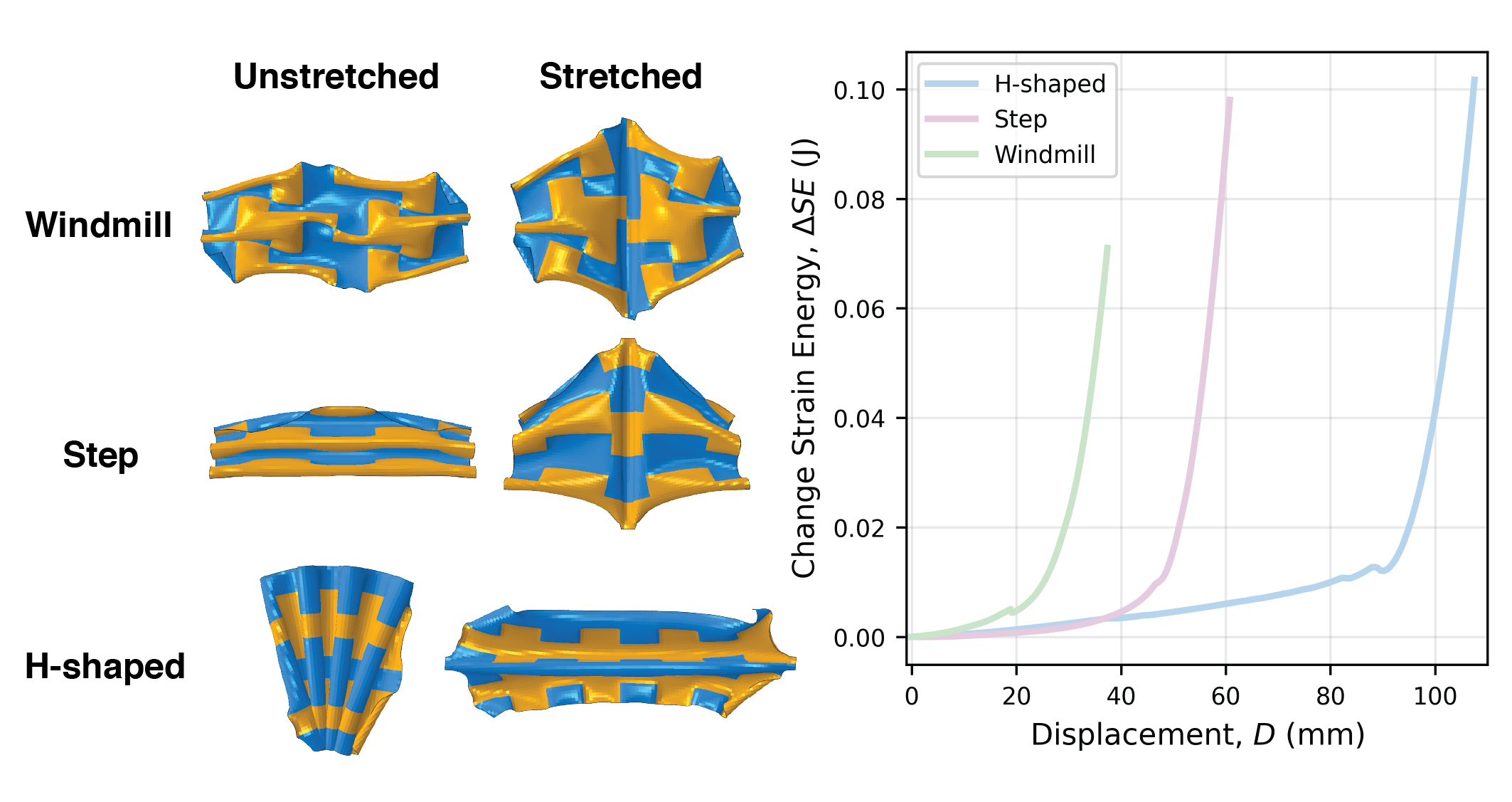}
  \caption{FEM strain energy results during point loading of three simulated geometries.}
  \label{si_fig:sim_results}
\end{figure}

Three rib-garter patterns were simulated as finite sized bi-layer equivalents for the H-shaped, Windmill, and Step patterns shown in Fig. \ref{fig:diff-geometries}. The strain-energy of each sample during loading is shown in Fig. \ref{si_fig:sim_results} alongside their configuration at the end of step 1 (labeled as unstretched) and step 2 (labeled as stretched). Critically, in the H-shaped model we see a clear strain energy minimum at a displacement of $\sim90$ mm, which corresponds to a second energy minimum in the sample and highlights the bilayers role in creating multistability. The same energy minimum is not observed in the Step sample. Finally, the Windmill model exhibits a shallow minimum at a displacement of approximately 20 mm.

\begin{figure}[h]
\centering
  \includegraphics[width=\linewidth]{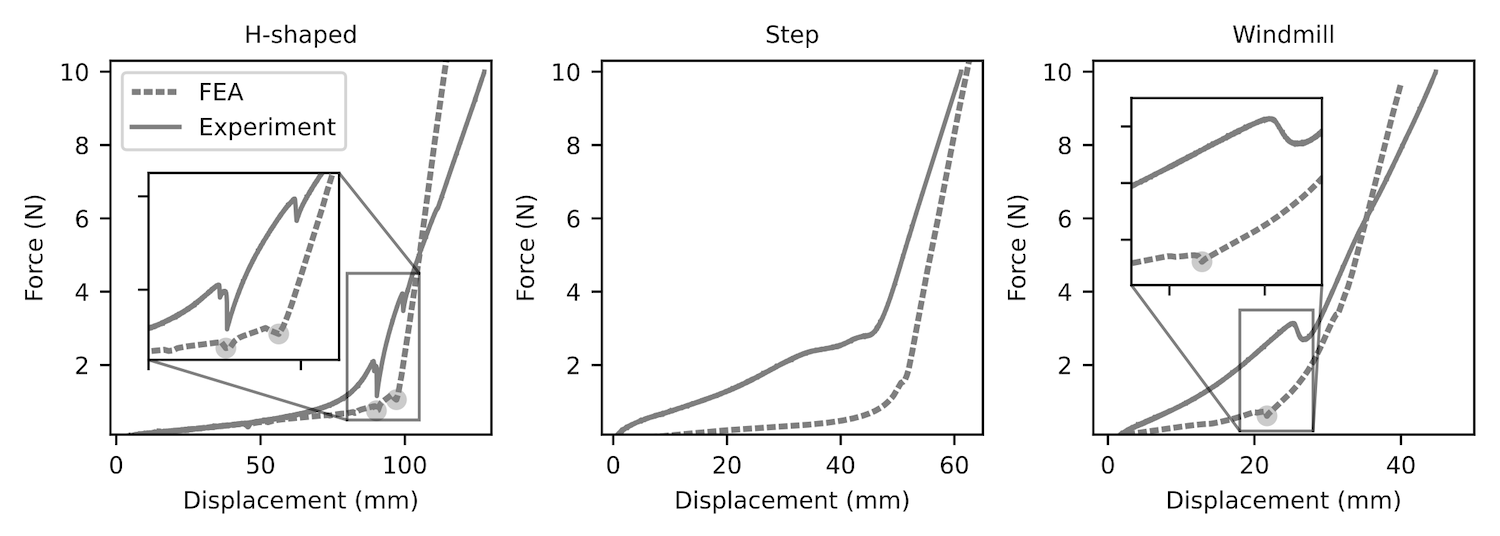}

  \caption{Comparison of the numerical predicted and experimentally measured   force-displacement curves for the H-shaped,  Step, and  Windmill patterns. We also include insets to show in detail the force-drop in the FE results.}
  \label{si_fig:force-disp-comparison}
\end{figure}

\new{
In Figure \ref{si_fig:force-disp-comparison} we compare the numerically predicted (dashed line) and experimentally measured (continuous lines)  force-displacement curves for the H-shaped, Step, and Windmill pattern. We find good agreement between the two sets of data. In full agreement with the experiments, the numerically predicted force-displacement curve  for the H-shape pattern  shows two force drops between 85 and 100 mm of displacement (highlighted by the grey circles) 
, after which the response stiffens as the shell becomes fully unfurled. For the Step pattern, both experimental and FE results show that there is no force drop, and stiffen at the sample displacement. Finally, both the FE and experimental results for the Windmill pattern show a force drop at similar displacements, and a subsequent stiffening. }


\new{In these simulations, we use a homogenized textile model to replicate the curvature and multistability of the knit samples through geometry and a pre-stressed bilayer. This type of homogenized model can help to understand system mechanics \cite{Farrell_2025,Connolly2016AutomaticMatching}, while avoiding the challenging and unresolved scaling complexity of yarn-level modeling \cite{ding2024unravelling}. These models are simplifications of reality and can never be perfectly accurate. For example, we predict that the yarn on yarn friction in the textiles is the cause of energy dissipation, but this dissipation is approximated with a finite element damping parameter. More accurate damping in simulation might lead to a better representation of the steep force drop. }

\newpage
\clearpage
\section{Additional Results}
~ 

\begin{figure}[h]
\centering
  \includegraphics{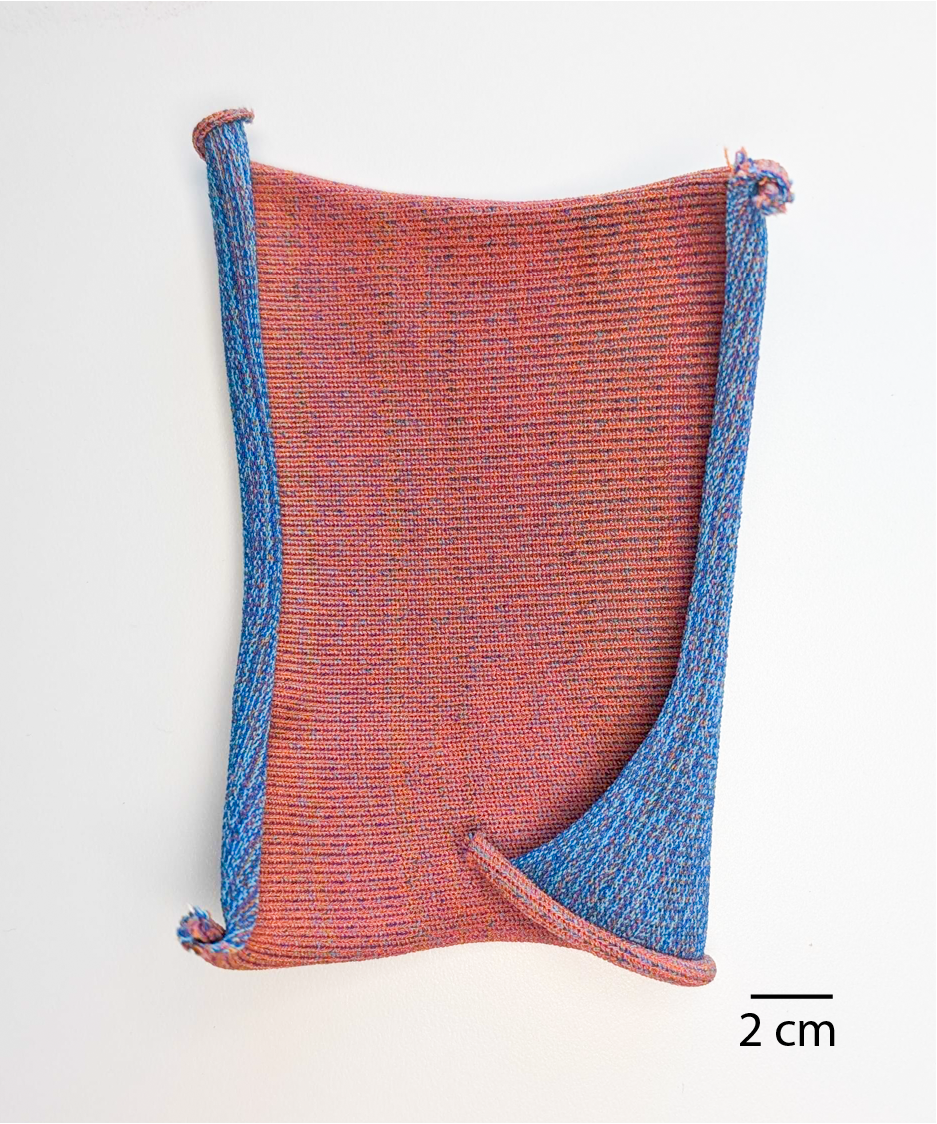}
  \caption{A larger sample of jersey knit fabric shows the curling concentrated at the edges. \textcolor{blue}{Sample is approximately 10cm wide.}}
  \label{si_fig:large_edge}
\end{figure}

\begin{figure}
\centering
  \includegraphics[width = \linewidth]{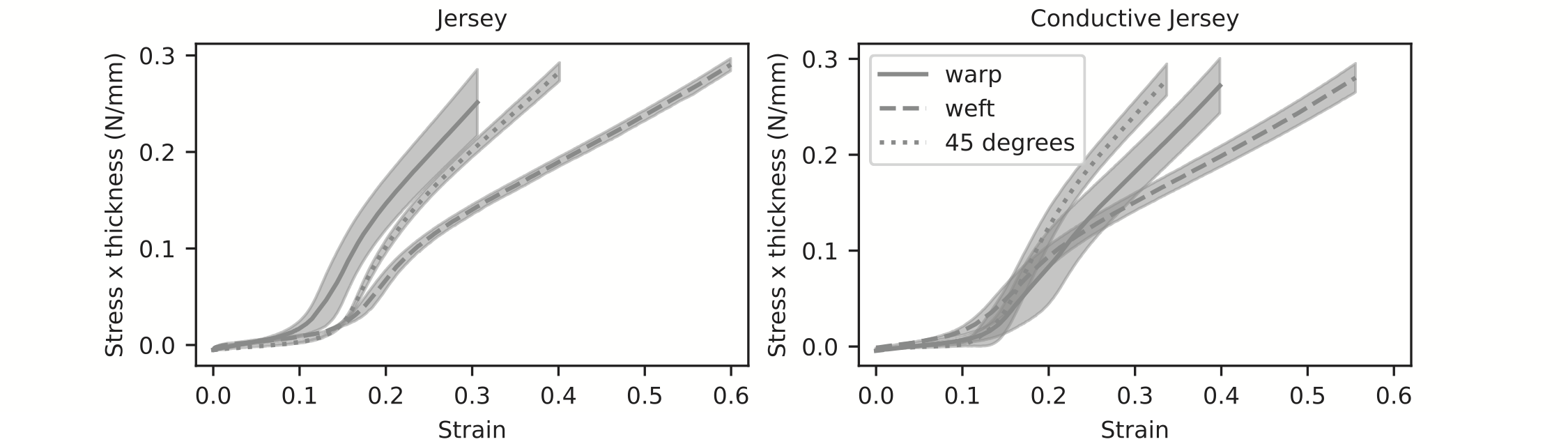}
  \caption{A) Mechanical response of a uniform, or jersey, fabric. Test samples were cut out of a larger textile so that the edges did not interfere. B) Mechanical response of a jersey fabric with two additional conductive yarns.}
  \label{si_fig:jersey}
\end{figure}
\begin{figure}[h]
  \centering\includegraphics[width = 0.4\linewidth]{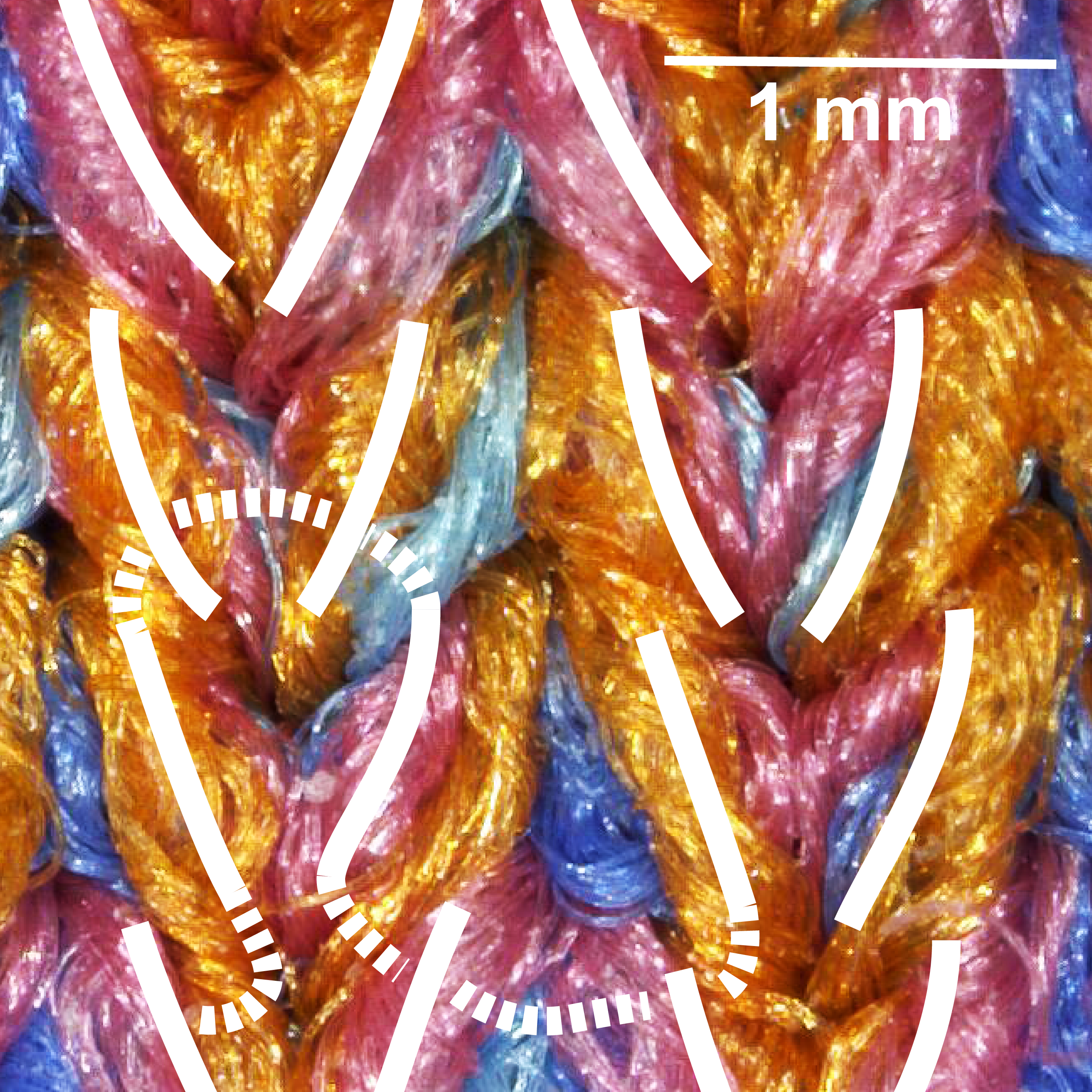}
  \includegraphics[width = 0.4\linewidth]{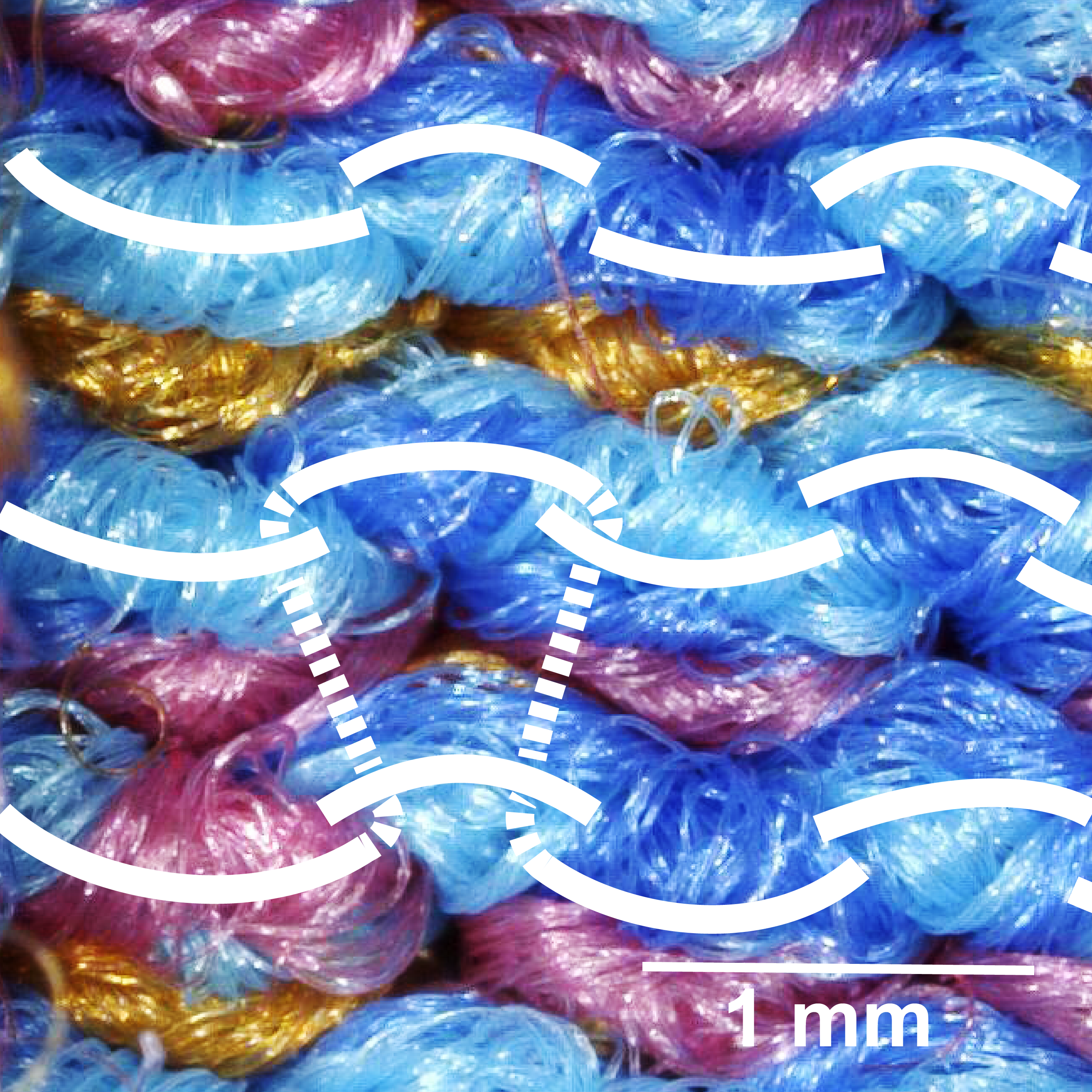}
  \caption{\new{Microscope images show the face and reverse of the knit fabric. The paths of the yarns are sketched on top, where the paths of the yarns that are on the opposite side are shown in dotted lines. This sketch shows us that the yarns on the  face lie mostly vertical and yarns on the reverse side lie mostly horizontal, with the crossovers happening through the thickness of the fabric.}}
  \label{si_fig:yarn_paths}
\end{figure}
\begin{figure}
\centering
  \includegraphics[width = 0.4\linewidth]{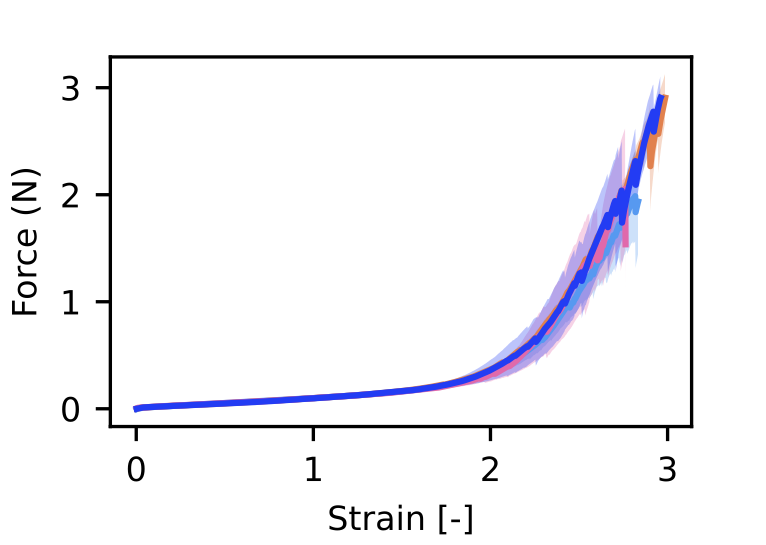}
  \caption{\new{Mechanical response of all four elastomeric yarns. We observe that the color does not affect the mechanical response}}
  \label{si_fig:elastomeric_yarns}
\end{figure}
\begin{figure}
\centering
  \includegraphics[width=\linewidth]{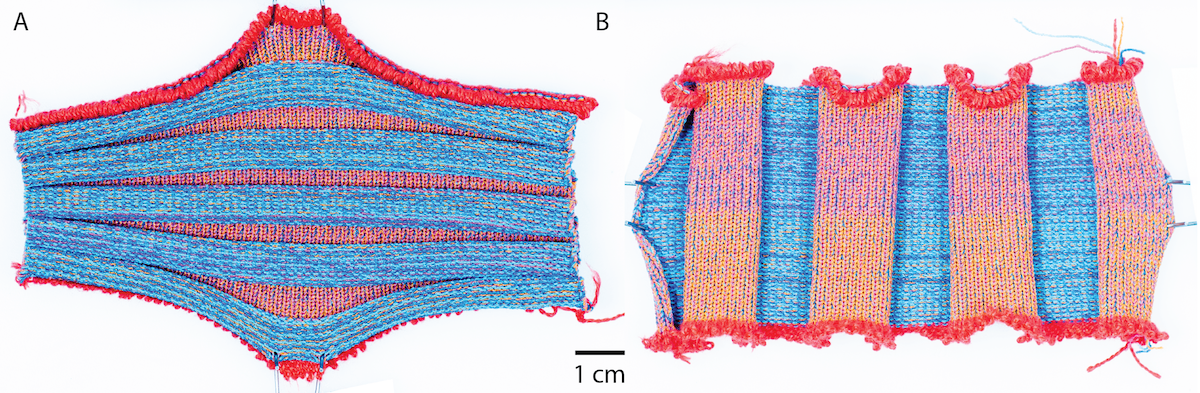}
  \caption{Rib and garter patterns are defined by the number of stitches in each stripe. Therefore a 12 x 12 pattern consists of 12 stitches knit on the front bed and 12 on the back bed. A) Top-down view of a 12 x 12 garter stitch, loaded at a single point to show both higher and lower strain regions. B) Top-down view of a 12 x 12 rib stitch, loaded at a single point to show both higher and lower strain regions. Top down views allow us to see the stripe patterns that define rib and garter knit patterns. Since each knit stitch is wider in the weft direction, the rib stripes are wider than those in the garter.}
  \label{si_fig:12 x 12 rib and garter}
\end{figure}
\begin{figure}
\centering
  \includegraphics[width=\linewidth]{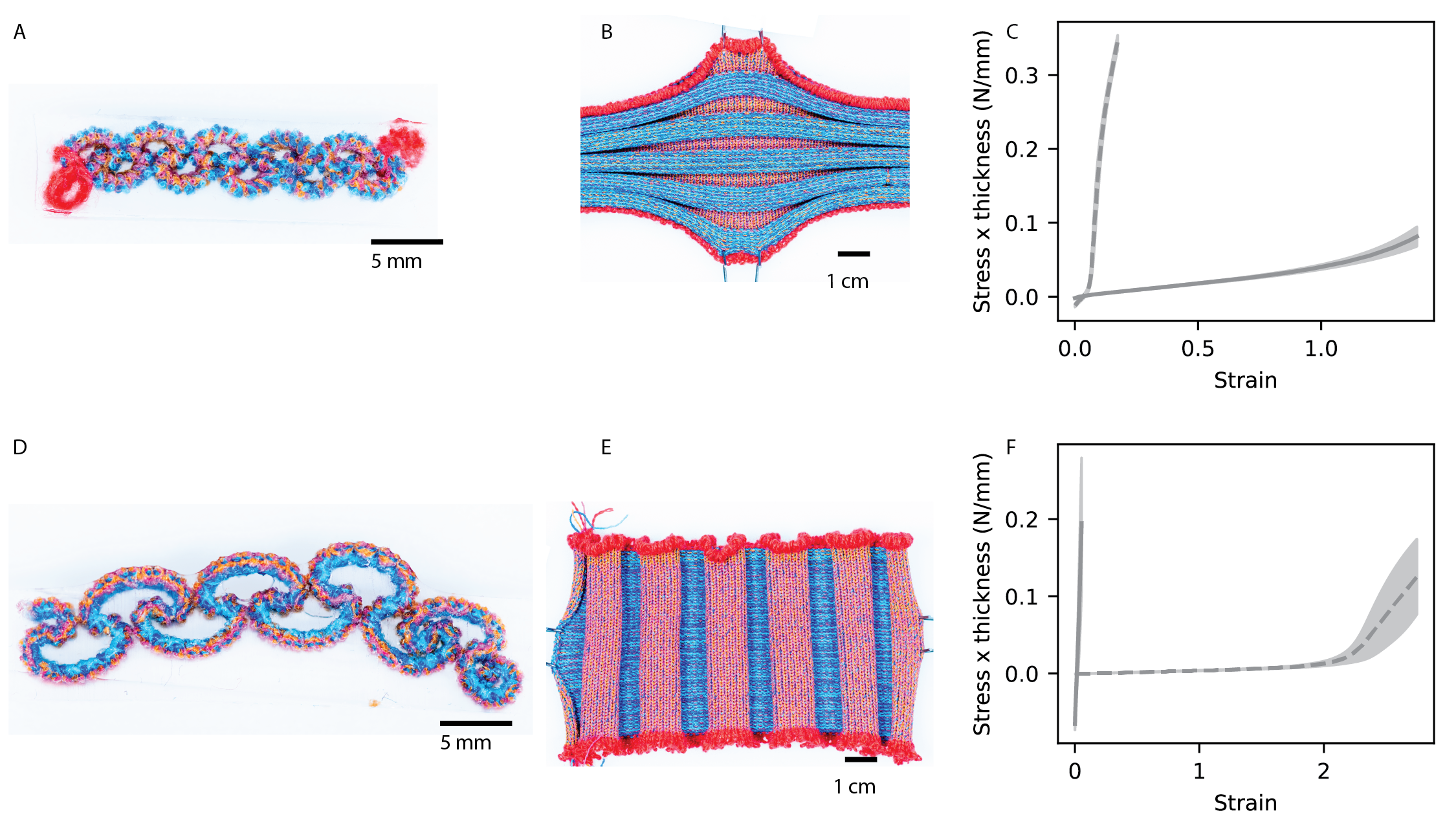}
  \caption{A) A cross section of an 8 x 8 garter pattern showing the corrugation and individual yarns. B) A top view of an 8 x 8 garter under a point load shows the textile under strain close to the center, and with no strain at the edges.  C) Stress-strain curves of 8 x 8 garter in the warp and weft direction  D) A cross section of an 8 x 8 rib pattern showing the corrugation and individual yarns  E) A top view of an 8 x 8 rib under a point load shows the textile under strain close to the center, and with no strain at the edges. F) Stress-strain curves of 8 x 8 rib in the warp and weft direction  }
  \label{si_fig:8 x 8 rib and garter}
\end{figure}

\begin{figure}
  \includegraphics[width=\linewidth]{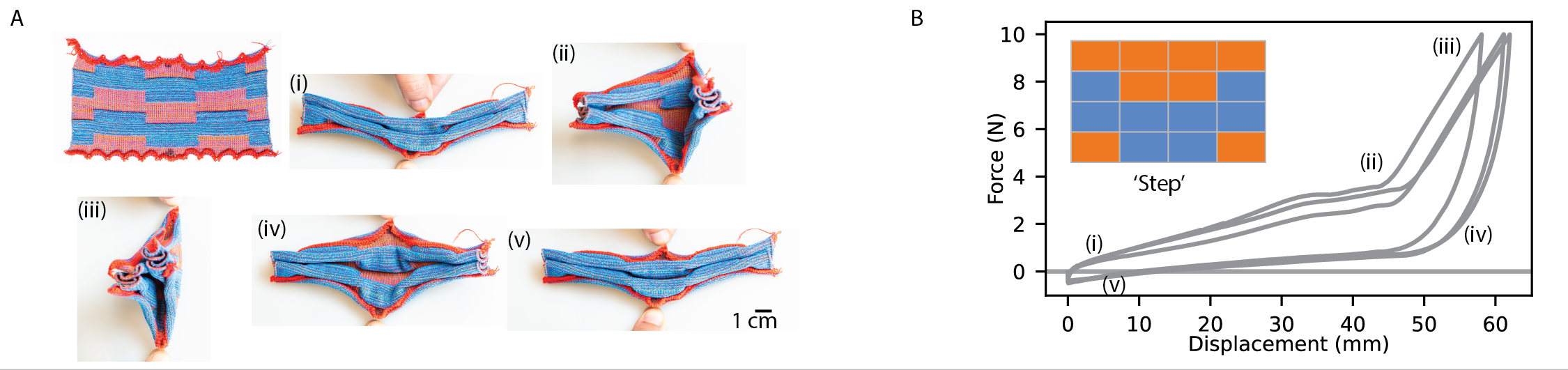}
  \caption{Additional photos and data for the \textit{Step} unit cell. A) Photos of the sample at various displacements, which are marked on B. B) Force-displacement response of sample. Photos at displacements (i) - (iv) are displayed in A}
  \label{si_fig:step}
\end{figure}

\begin{figure}
\centering
  \includegraphics[width=\linewidth]{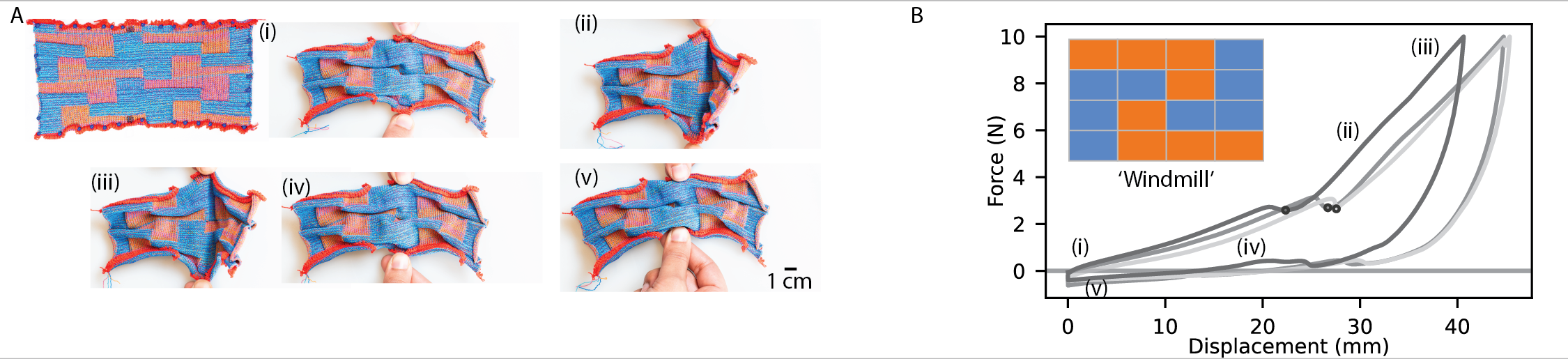}
  \caption{Additional photos and data for the \textit{Windmill} unit cell. A) Photos of the sample at various displacements, which are marked on B. B) Force-displacement response of sample. Photos at displacements (i) - (iv) are displayed in A}
  \label{si_fig:windmill}
\end{figure}

\begin{figure}
\centering
  \includegraphics[width=\linewidth]{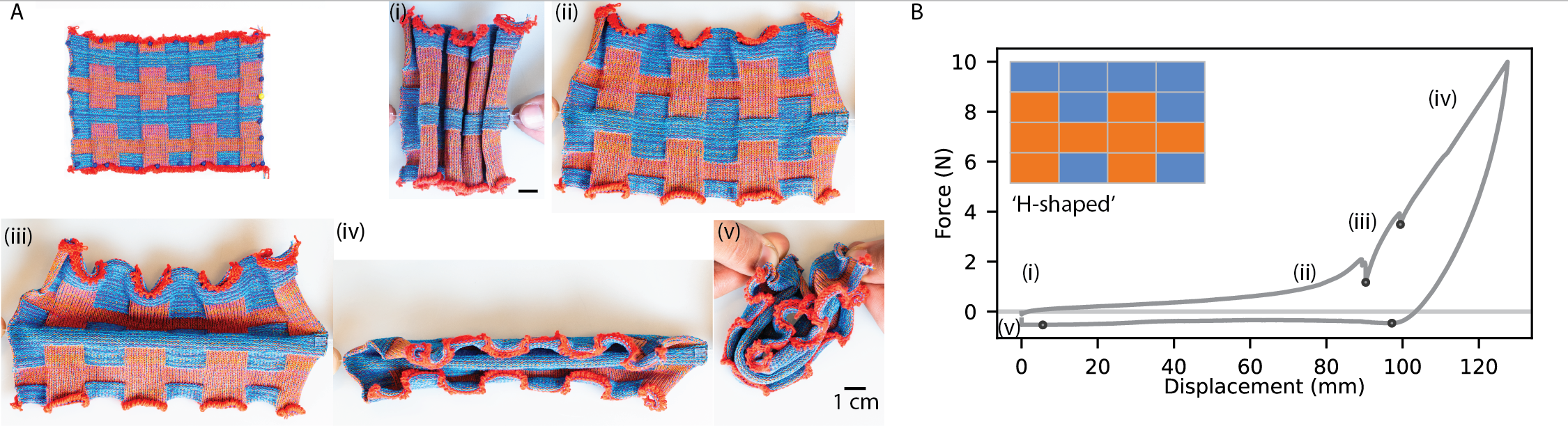}
  \caption{A) Additional photos for the \textit{H-shaped} unit cell. A) Photos of the sample at various displacements, which are marked on B. B) Force-displacement response of sample. Photos at displacements (i) - (iv) are displayed in A. Further force-displacement data is shown in \ref{si_fig:d12 graphs}.}
  \label{si_fig:h-shaped}
\end{figure}

\begin{figure}
\centering
  \includegraphics[width=\linewidth]{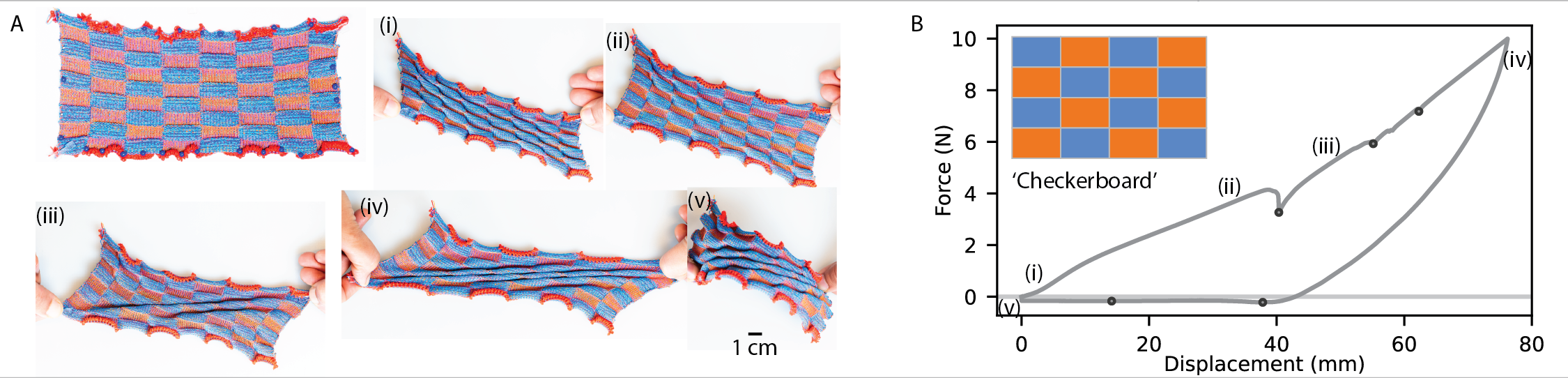}
  \caption{Photos and data for the \textit{Checkerboard} unit cell. This knit pattern displays an unusual pattern of diagonal folds, which can fold either tilted left or right. The two possible fold directions allow for multistability. A) Photos of the sample at various displacements, which are marked on B. B) The force-displacement response for the checkerboard pattern when it is loaded along the diagonal direction. Photos at displacements (i) - (iv) are displayed in A}
  \label{si_fig:checkerboard}
\end{figure}

\begin{figure}
\centering
  \includegraphics[width=\linewidth]{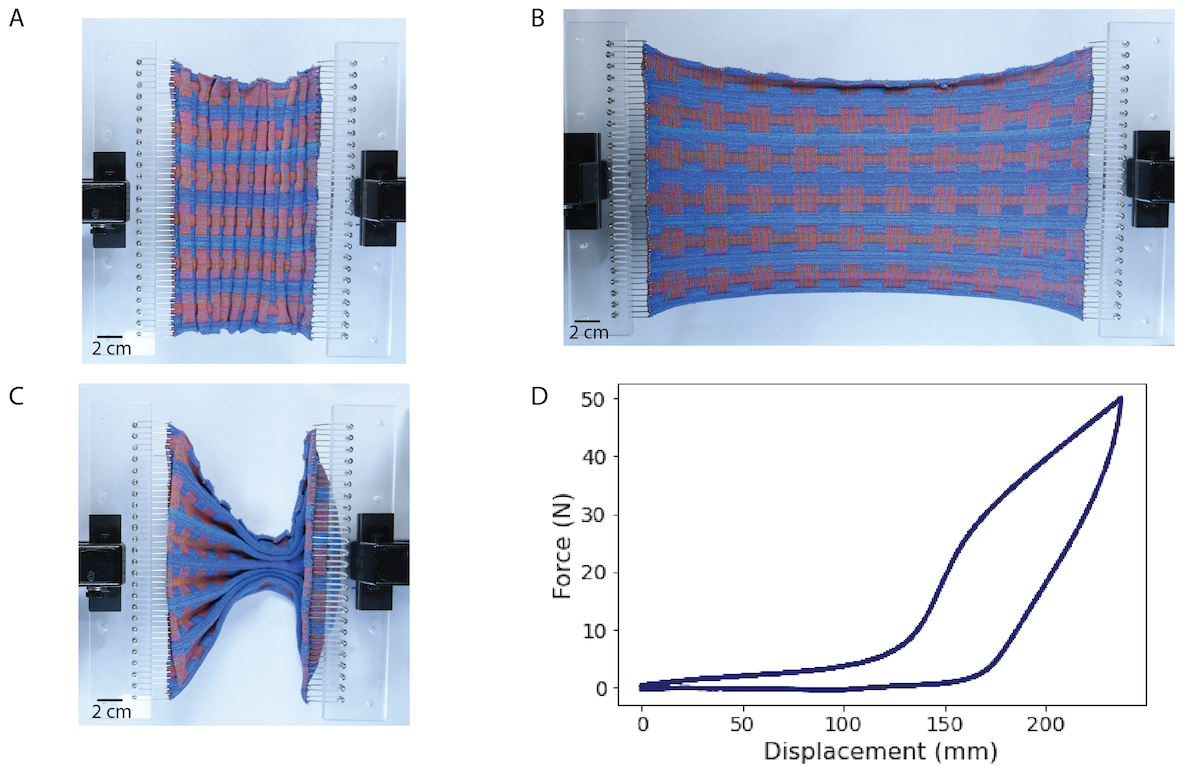}
  \caption{A) A larger sample of \emph{`H-shaped'} unit cells is clamped along the edges. B) Strain is applied to the sample. C) When unloaded, the unit cells in the center of the sample have switched configurations. D) The force-displacement response of this clamped sample.}
  \label{si_fig:clamped}
\end{figure}

\begin{figure}
\centering
  \includegraphics[width=\linewidth]{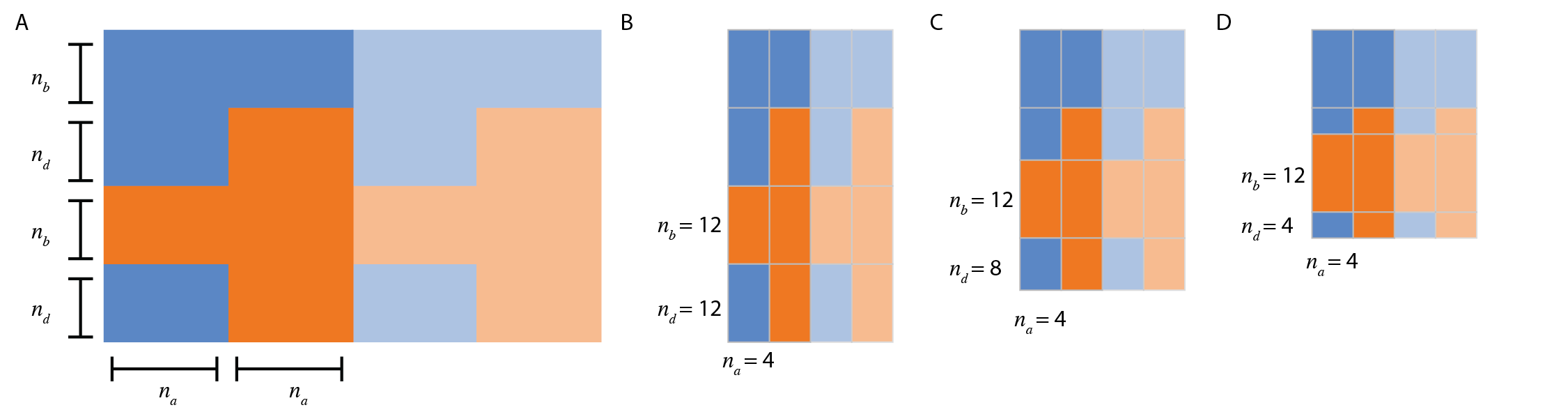}
  \caption{A) The 4 x 4 unit cell is made up of two identical 4 x 2 sections. This unit cell can be characterized by six parameters, which we reduce to three to maintain symmetry. B) A schematic of the sample shown in Fig. \ref{fig:H-shape}C-i. C) ) A schematic of the sample shown in Fig. \ref{fig:H-shape}C-ii. ) A schematic of the sample shown in Fig. \ref{fig:H-shape}C-iii.}
  \label{si_fig:hunits}
\end{figure}

\begin{figure}[h]
\centering
  \includegraphics[width=\linewidth]{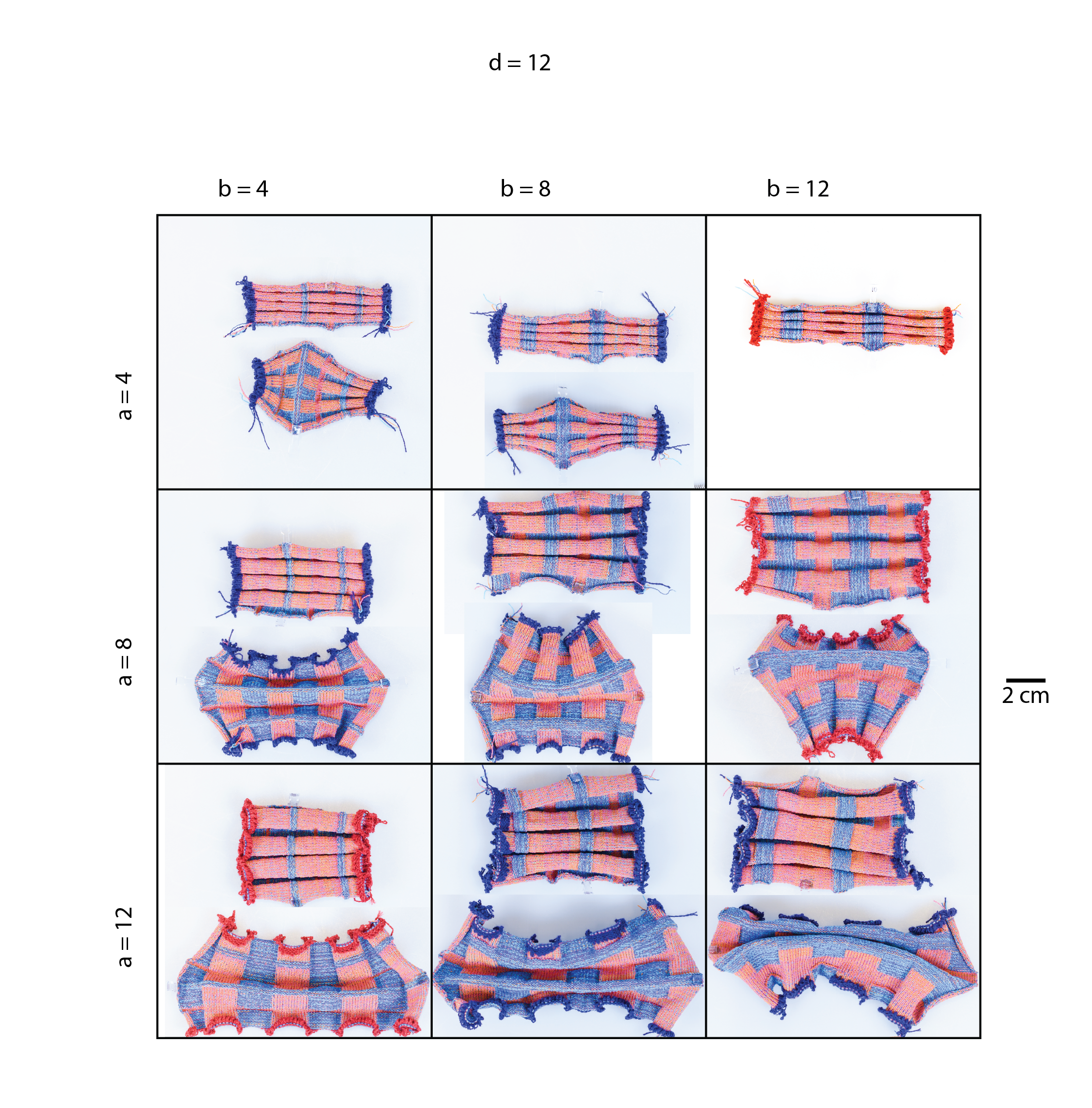}
  \caption{Photos of samples with \textit{H-shaped} unit cells where $n_d = 12$.}
  \label{si_fig:d12 photos}
\end{figure}

\begin{figure}[h]
\centering
  \includegraphics[width=\linewidth]{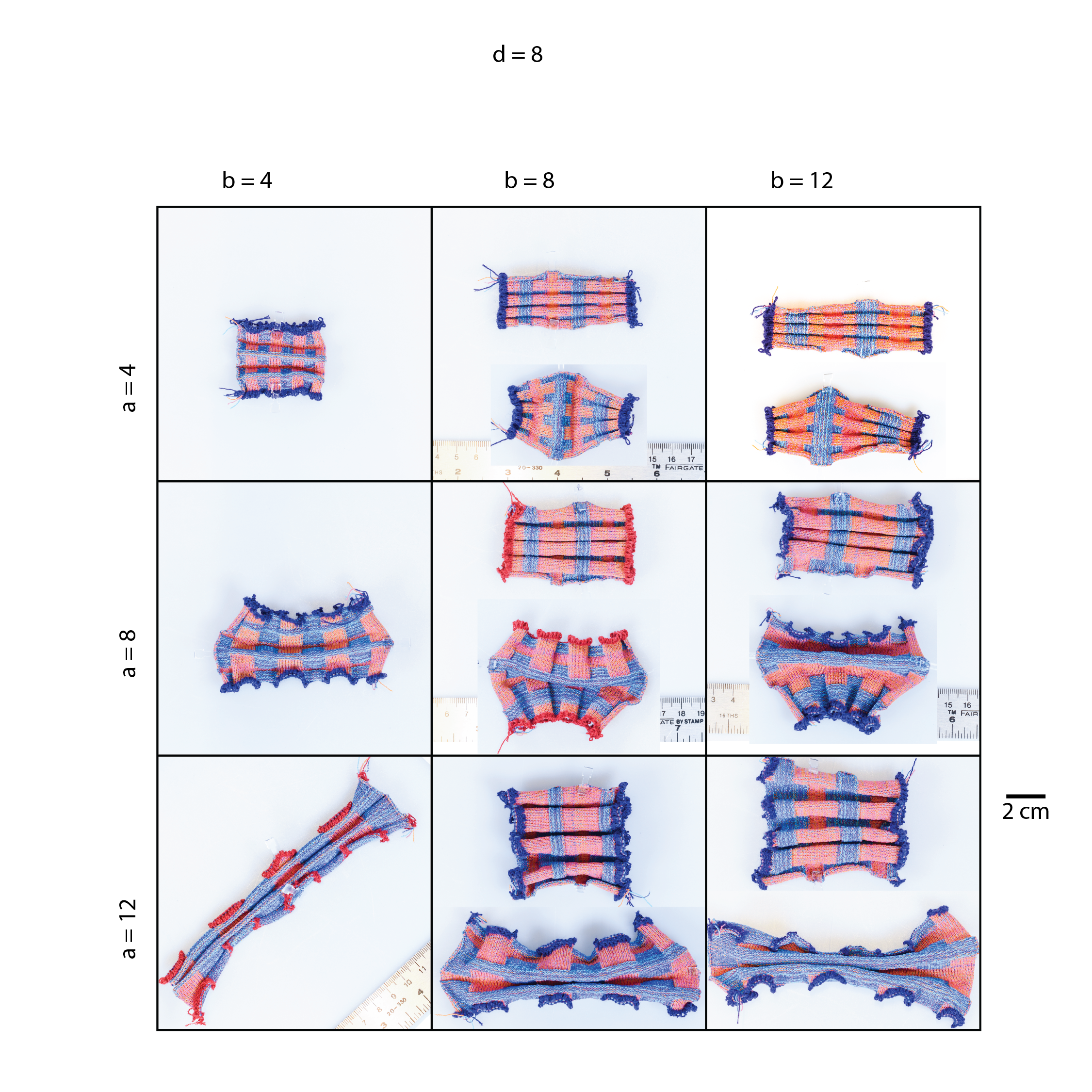}
  \caption{Photos of samples with \textit{H-shaped} unit cells where $n_d = 8$.}
  \label{si_fig:d8 photos}
\end{figure}

\begin{figure}[h]
\centering
  \includegraphics[width=\linewidth]{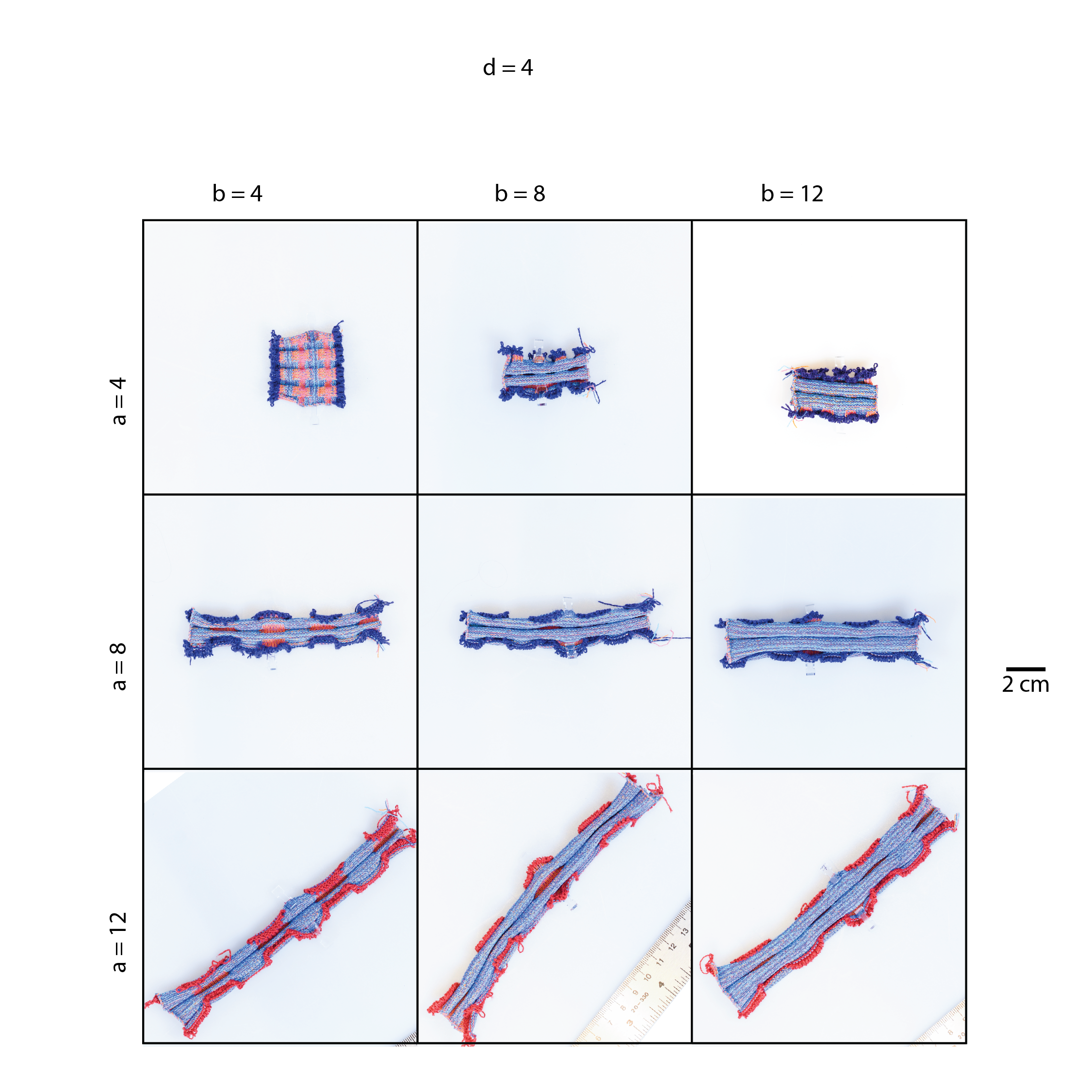}
  \caption{Photos of samples with \textit{H-shaped} unit cells where $n_d = 4$.}
  \label{si_fig:d4 photos}
\end{figure}

\begin{figure}[h]
\centering
  \includegraphics[width=\linewidth]{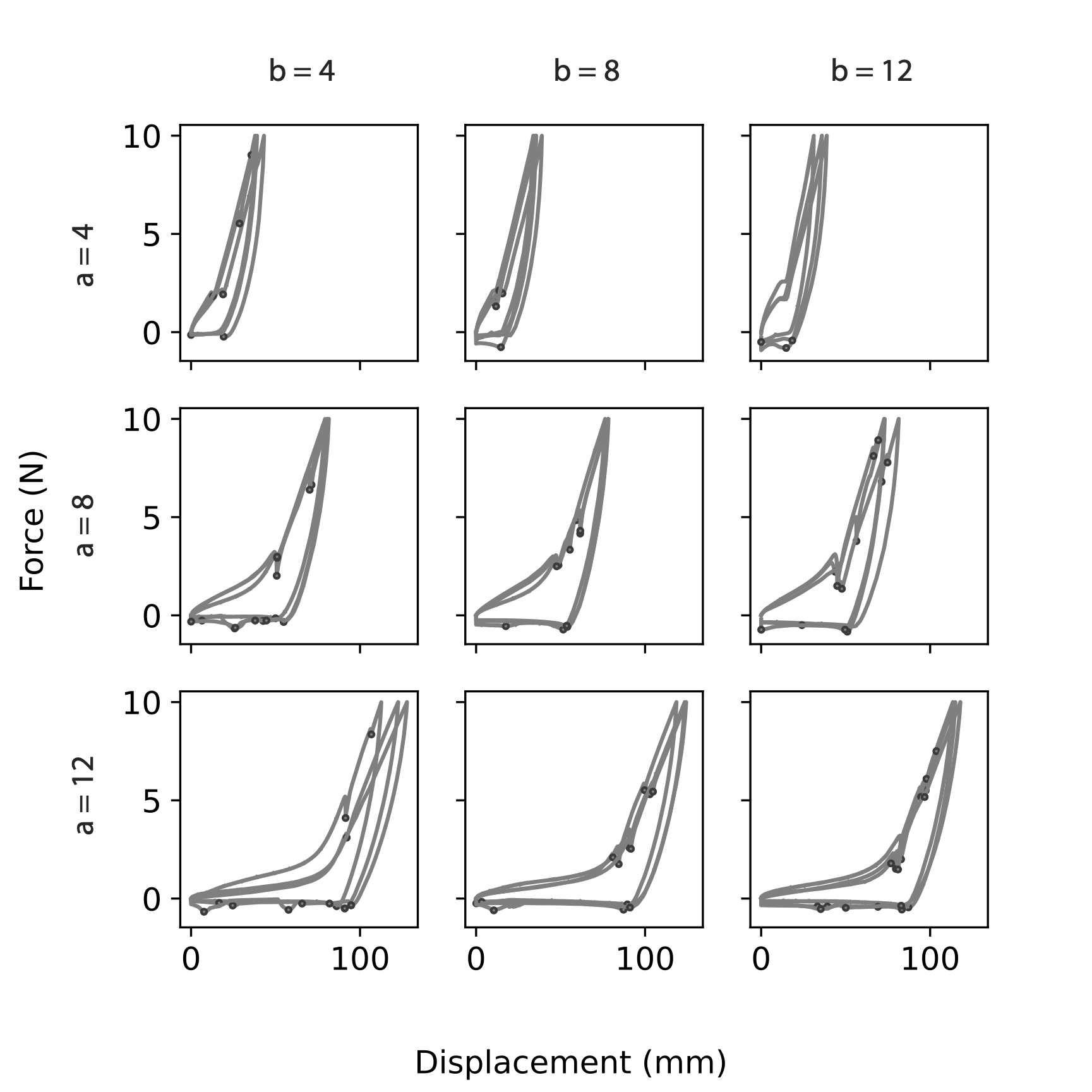}
  \caption{Additional data of samples with \textit{H-shaped} unit cells where $n_d = 12$.}
  \label{si_fig:d12 graphs}
\end{figure}

\begin{figure}[h]
\centering
  \includegraphics[width=\linewidth]{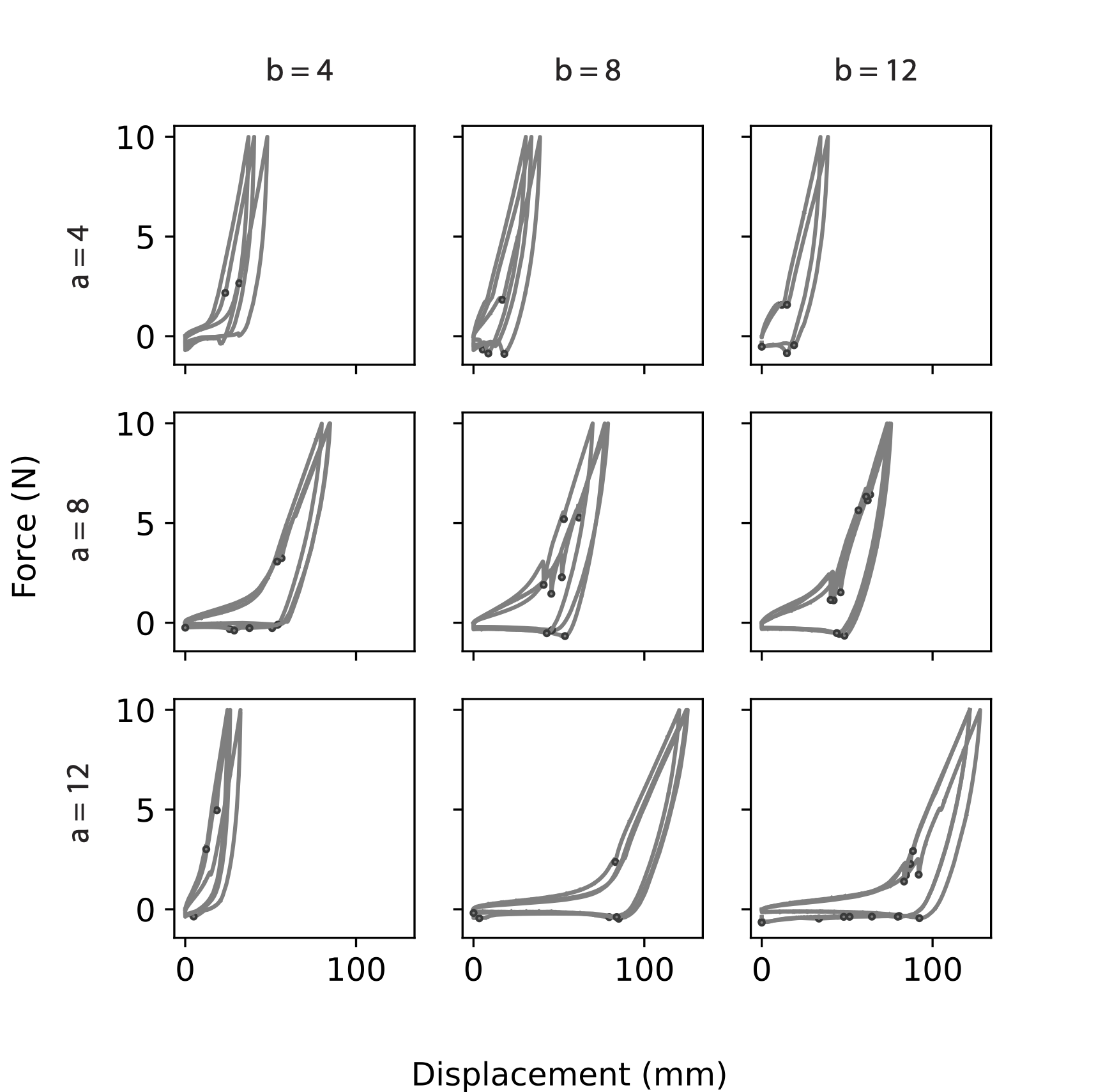}
  \caption{Additional data of samples with \textit{H-shaped} unit cells where $n_d = 8$.}
  \label{si_fig:d8 graphs}
\end{figure}

\begin{figure}[h]
\centering
  \includegraphics[width=\linewidth]{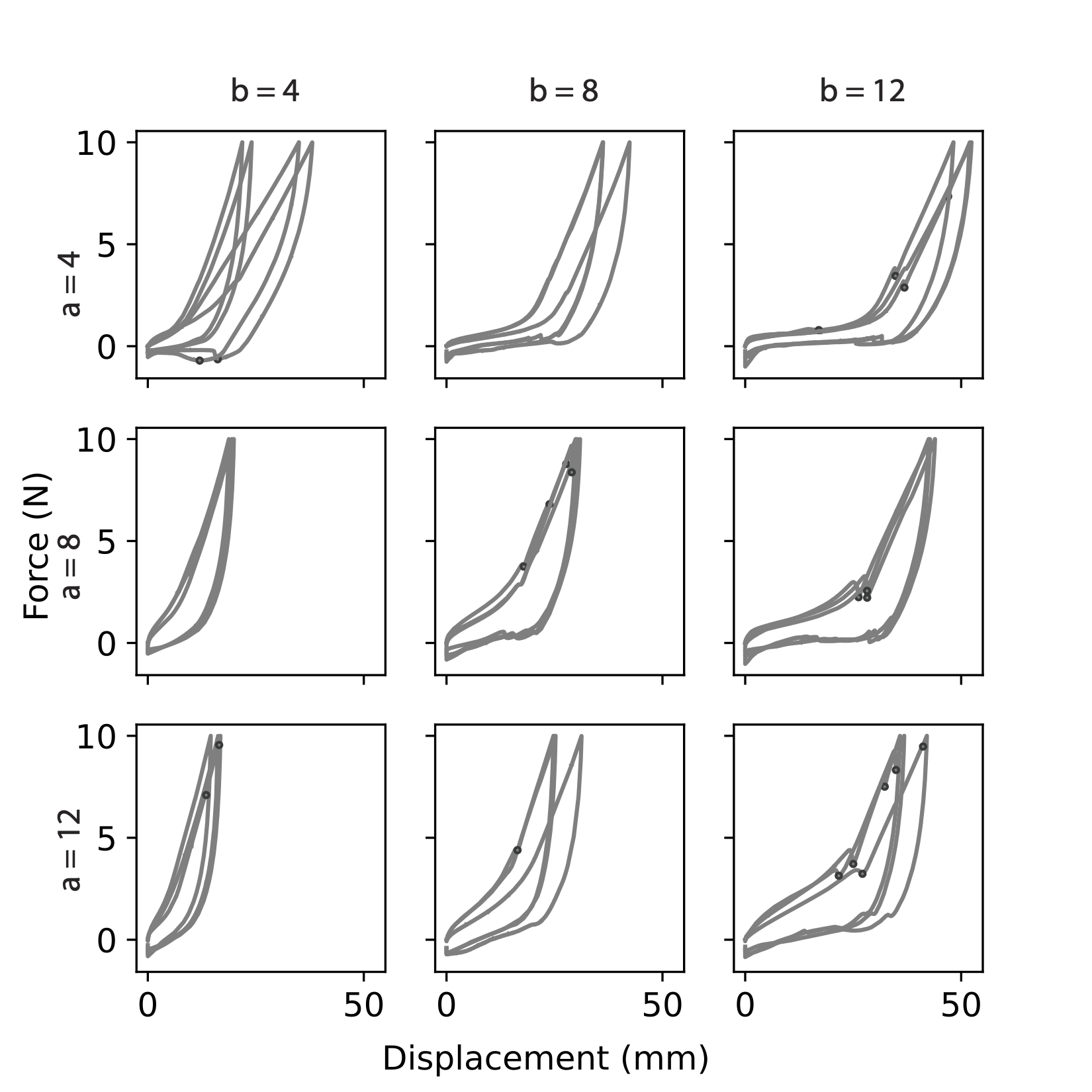}
  \caption{Additional data of samples with \textit{H-shaped} unit cells where $n_d = 4$.}
  \label{si_fig:d4 graphs}
\end{figure}

\begin{figure}[h]
  \centering\includegraphics{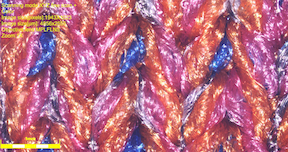}
  \includegraphics{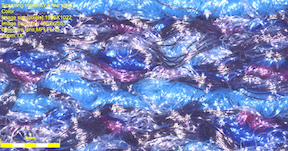}
  \caption{Microscope images show the front and back of the knit fabric which includes conductive yarns. The silver-plated yarns are visible between the orange and blue yarns. Scale bars are 1 mm.}
  \label{si_fig:conductive fabric}
\end{figure}
\clearpage

\newpage
\clearpage
\section{Movies}
\noindent \textbf{Movie S1:} Knit structures under tensile loading\\
\noindent \textbf{Movie S2:} Knit switches with built-in haptic feedback\\
\noindent \textbf{Movie S3:} Applications of a momentary switch\\
\noindent \textbf{Movie S4:} Reconfigurable latching switch lamp

\newpage
\clearpage
\printbibliography[title={References}]
\end{refsection}
\end{document}